\documentclass{article} 

\usepackage{amsmath,amsthm} 
\usepackage{amssymb,mathrsfs} 
\usepackage{a4wide} 
\usepackage{hyperref}
\usepackage{graphicx}
\usepackage{color,subfigure} 
\usepackage{enumitem}
\usepackage{fullpage}
\usepackage[pdftex,dvipsnames]{xcolor}  
\usepackage{xargs}                      
\usepackage[normalem]{ulem}

\newtheorem{theorem}{Theorem}[section]

\newtheorem{remark}[theorem]{Remark}

\newcommand{\rmo}{\mathrm{o}}
\newcommand{\rme}{\mathrm{e}}
\newcommand\RR{{\mathbb R}}
\renewcommand{\bar}{\overline}
\renewcommand{\le}{\leqslant}
\renewcommand{\leq}{\leqslant}
\renewcommand{\ge}{\geqslant}
\renewcommand{\geq}{\geqslant}
\newcommand{\dps}{\displaystyle}
\newcommand{\rL}{\mathrm{L}}
\newcommand{\rR}{\mathrm{R}}
\newcommand{\TT}{\mathbb{T}}
\renewcommand{\ss}{\mathrm{ss}}
\newcommand{\eps}{\varepsilon}

\newcommand\ds{\displaystyle}
\newcommand*\dd{\mathop{}\!\mathrm{d}}
\renewcommand{\S}{S}

\numberwithin{equation}{section}

\begin{document}

\title{Thermo-mechanical transport in rotor chains}
\author{A. Iacobucci$^{1}$, S. Olla$^{1}$ and G. Stoltz$^{2,3}$ \\
	{\small $^{1}$ CNRS, CEREMADE, Universit\'e Paris-Dauphine, PSL Research University, 75016 Paris, France} \\
	{\small $^{2}$ CERMICS, Ecole des Ponts, Marne-la-Vall\'ee, France} \\
        {\small $^{3}$ MATHERIALS team-project, Inria Paris, France} \\
}

\date{\today} 

\maketitle

\begin{abstract}
We study the macroscopic profiles of temperature and angular momentum in the stationary state
of chains of rotors under a thermo-mechanical forcing applied at the boundaries.
These profiles are solutions of a system of diffusive partial differential
equations with boundary conditions determined by the thermo-mechanical forcing. 
Instead of expensive Monte Carlo simulations of the underlying microscopic dynamics,
we perform extensive numerical computations based on a finite difference method for
the system of partial differential equations describing the macroscopic steady state. 
We first present a formal derivation of these stationary equations based on a linear response argument
and local equilibrium assumptions. We then study various properties of the solutions to these equations.
This allows to characterize the regime of parameters leading to uphill energy diffusion --
a situation in which the energy flows in the direction of the gradient of temperature --
and to identify regions of parameters corresponding to a \emph{negative energy conductivity} (\emph{i.e.} a positive linear response of the energy current to a gradient of temperature).
The macroscopic equations we derive are consistent with some previous results obtained by numerical simulation of the microscopic physical system, which confirms their validity.
\end{abstract}

\section{Introduction}
\label{intro}

A rigorous understanding of the microscopic origin of Fourier's law is still elusive,
despite the progress achieved by means of numerous scientific works in the last decades (see~\cite{blr} for a review which is still up-to-date in many aspects,
as well as~\cite{BF19} for contemporary perspectives).
Fourier's law claims that the local energy current is proportional to the opposite of the local gradient of temperature.
The ratio of these quantities in absolute value is a function of the local temperature and is called thermal conductivity.
A particular subclass of models has attracted attention in the mathematical
and theoretical physics literature on thermal transport
already at the end of the 1990s, namely one dimensional chains of atoms
(see the review articles~\cite{LLP03,Dhar08,ILLPP19}).
The idea was to consider the simplest possible model to understand
the sufficient and/or necessary ingredients for Fourier's law to hold.
This proved to be a challenging problem, since the thermal conductivity of one dimensional chains is usually anomalous, \emph{i.e.} it does not admit a well defined thermodynamic limit for increasing system size. 
More recently, owing in particular to the carbon nanotube industry, it turned out that these caricatural one dimensional systems are of the uttermost physical relevance, and that the thermal conductivity divergence predicted by numerical simulations can actually be
observed for sufficiently clean experimental samples~\cite{review_exp}.

We consider a chain of coupled rotors, which is one of the simplest one dimensional models for which Fourier's law holds.
Various authors studied numerically the properties of the non-equilibrium stationary state obtained by applying
Langevin thermostats at different temperatures at the boundaries~\cite{GLPV00,GS00,YH05,GS05}.
In particular, they found that the thermal conductivity (defined as the absolute value of the stationary energy current multiplied by the system size and divided by the boundary temperature difference) has a finite limit for large system sizes. Computations by Green-Kubo formulas give the same results, at least for temperatures not too low.
One can further characterize the spreading of space-time correlations,
relying on the theory of nonlinear hydrodynamics~\cite{Spohn14,DD14}. 

However, in many realistic systems, there are other conserved quantities besides energy, and the interplay among them has a deep impact on the system thermal properties, in particular when all these conserved quantities evolve on the macroscopic diffusive scale~\cite{Olla19}.
In the specific case of one dimensional rotor chains, the extra conserved quantity is angular momentum.
As a consequence, the physical behavior of the stationary state becomes much more
interesting when a mechanical forcing is applied at the boundaries in addition to the thermal forcing.
The mechanical forcing creates a current of angular momentum,
and the local momentum gradients interact with those of energy,
giving rise to highly non trivial stationary profiles.

In~\cite{ILOS11}, we performed numerical simulations of rotor chains under thermo-mechanical forcing and observed some intriguing physical phenomena, such as the appearance of nonmonotonic temperature profiles in the steady state and a negative energy conductivity in some regimes of parameters
--- by which we mean that the energy current magnitude increases (resp. decreases)
as the absolute value of the temperature difference is decreased (resp. increased).
Yet, it would be computationally expensive to thoroughly analyze the origin and implications of these fascinating phenomena by microscopic simulations, since one should consider
sufficiently large systems and integrate their dynamics over appropriately long times. A more affordable alternative is to rely on the macroscopic description of the behavior of the conserved physical quantities associated to the microscopic dynamics.

In order to derive this macroscopic description, we consider the stationary case (see~\cite{Olla19} for the time-dependent problem under a space-time diffusive scaling), and rely on a local equilibrium assumption, that has been numerically verified in~\cite{ILOS11}. It is then possible to associate 
stationary profiles of temperature to stationary profiles of the conserved quantities
(energy and angular momentum). These profiles must 
satisfy the diffusive system~\eqref{eq:19}, which could also be deduced from a generic
argument proposed in~\cite{ILLP16}.
We present here a formal derivation of~\eqref{eq:19},
based on a linear response argument, under a local equilibrium assumption and taking into account first order corrections.
A mathematically rigorous proof of such a derivation is a very
challenging open problem. It has been performed for a much simpler dynamics
with two conserved quantities and constant transport coefficients~\cite{kos19,kos20}.
Various symmetries and mathematical properties of the stationary solution can be directly deduced from~\eqref{eq:19} and are precisely discussed in this work.

After computing the transport coefficients by means of equilibrium microscopic simulations, we {nu\-me\-ri\-cal\-ly} solve the stationary equations~\eqref{eq:19}.
The agreement between the results presented here and those obtained by microscopic dynamics simulation in~\cite{ILOS11} supports the validity of the system~\eqref{eq:19}, formally derived thorough linear response and local equilibrium arguments. We then carefully study the phenomenon of uphill energy diffusion, which appears when the usual heat diffusion is counterbalanced by an energy current induced by the mechanical forcing. Uphill diffusion was also proved in a simple model in~\cite{kos20}, and observed for other particle systems such as two-dimensional Ising systems~\cite{CGGV18}, one dimensional discrete nonlinear Schrodinger chains~\cite{ILLOP17} and one-dimensional Hamiltonian systems~\cite{WCB20}. We also determine ranges of parameters leading to the appearance of negative energy conductivity.

\paragraph{Outline of the work.} We derive the system of partial differential equations satisfied by the stationary profiles in Section~\ref{sec:theory} and discuss the numerical results obtained by solving this system in Section~\ref{sec:numerics}. Details on the methods used to generate the numerical results are provided in Appendices~\ref{App:GKnum} and~\ref{App:SSnum}.

\section{Derivation and properties of the effective diffusion system}
\label{sec:theory}

We present the atomistic dynamics (Section~\ref{ssec:microdyn}) and define the associated linear response coefficients (Section~\ref{ssec:onsager-matrix}). We then derive the system of effective equations formally describing the stationary state of the system in the macroscopic limit ( Section~\ref{ssec:stationary-profiles}). We gather various analytical properties of this stationary state (Section~\ref{ssec:ssproperties}), and conclude with a qualitative discussion of the phenomena arising from thermo-mechanical forcing (Section~\ref{sec:qualitative_phenomena}).

\subsection{Microscopic dynamics}
\label{ssec:microdyn}

We consider a chain of $2N+1$ particles, described by its configuration~$(q_{-N},\dots,q_N,p_{-N},\dots,p_N) \in \TT^{2N+1} \times \RR^{2N+1}$, where each position~$q_i$ is defined on the one-dimensional torus $\TT = \RR\backslash(2\pi\mathbb{Z})$, with free boundary conditions. At the left and right boundaries, we add two constant forces of magnitudes~$\tau_\rL,\tau_\rR \in \RR$  and two Langevin thermostats at temperatures~$T_\rL,T_\rR$. Denoting the relative angle between rotors by $r_i = q_i - q_{i-1}$ for $i= -N+1, \dots, N$, the Hamiltonian of the system is
\[
  H(q,p) = \sum_{i=-N}^N e_i, \qquad e_i = \frac{p_i^2}{2} + V(r_i) \textrm{ for }
  i=-N+1,\dots,N, \qquad e_{-N} = \frac{p_{-N}^2}{2},
\]
with $V(r) = 1-\cos(r)$.

\paragraph{Dynamics.} The dynamics of the system in the bulk region reads
\begin{equation}
\label{eq:dyn_bulk}
\begin{aligned}
  \dot r_i(t) &= p_i(t) - p_{i-1}(t) , & \qquad i & = - N+1, \dots, N,\\
  \dot p_i (t)&= V'(r_{i+1}(t)) -  V'(r_{i}(t)), & \qquad i & = - N+1, \dots, N-1, \\
\end{aligned}
\end{equation}
while, at the boundaries,
\begin{equation}
  \label{eq:dyn_boundary}
  \begin{aligned}
    \dd p_{-N} (t) &= \left(\tau_\rL + V'(r_{-N+1}(t)) - \gamma p_{-N}(t)\right) \dd t + \sqrt{2\gamma T_\rL} \, \dd W_\rL(t),\\
    \dd p_N (t) &= \left(\tau_\rR -  V'(r_{N}(t)) - \gamma p_N(t) \right) \dd t + \sqrt{2\gamma T_\rR} \, \dd W_\rR(t),
  \end{aligned}
\end{equation}
where $W_\rL,W_\rR$ are two independent standard Brownian motions. The generator of the process can be written as
\[
  L_{N,T_\rL,T_\rR,p_\rL,p_\rR} = A_N + \gamma \left(\S_{\rL} + \S_{\rR}\right),
\]
with a Hamiltonian part
\[
  A_N = \sum_{i=-N+1}^N \left\{\left(p_i - p_{i-1}\right) \partial_{r_i}- V'(r_i) \left(\partial_{p_{i}} - \partial_{p_{i-1}}\right) \right\},
\]
and the generators of the Ornstein--Uhlenbeck processes at the boundaries
\[
  \S_\rL = T_\rL \partial_{p_{-N}}^2 - (p_{-N} - p_\rL) \partial_{p_{-N}} ,
  \qquad
  \S_\rR  = T_\rR \partial_{p_{N}}^2 - (p_{N} - p_\rR) \partial_{p_{N}},
\]
where $p_\rL = \gamma^{-1} \tau_\rL$ and $p_\rR = \gamma^{-1} \tau_\rR$.

\paragraph{Currents.} There are two locally conserved (or balanced) quantities:
the momentum~$p_i$ and the energy~$e_i$.
In fact, denoting by $J^a_{i,i+1}(t)$ (with $a \in \{p,e\}$) the corresponding total currents up to time $t$,
the following conservation laws hold for $i= -N, \dots, N-1$: for the momentum,
\begin{equation}
  \label{eq:1a}
  \dd p_i(t) = \dd J^p_{i-1,i}(t) -\dd J^p_{i,i+1}(t), \qquad \dd J^p_{i,i+1}(t) =  j^p_{i,i+1}(t) \, \dd t = -V'(r_{i+1}(t)) \, \dd t,
\end{equation}
with boundary currents
    \[
      \dd J^p_{-N-1, -N}(t) = \left(\tau_\rL - \gamma p_{-N}\right) \dd t + \sqrt{2\gamma T_\rL} \, \dd W_\rL(t),
      \qquad \dd J^p_{N, N+1}(t) = \left(-\tau_\rR +\gamma p_{N}\right) \dd t - \sqrt{2\gamma T_\rR} \, \dd W_\rR(t);
    \]
while, for the energy, 
\begin{equation}
  \label{eq:1b}
  \dd e_i(t) = \dd J^e_{i-1,i}(t) - \dd J^e_{i,i+1}(t), \qquad
  \dd J^e_{i,i+1}(t) = j^e_{i,i+1}(t) \, \dd t = -p_i(t) V'(r_{i+1}(t)) \, \dd t, 
\end{equation}
with boundary currents 
\[
  \begin{aligned}
    \dd J^e_{-N-1, -N}(t) & = \left(\tau_\rL p_{-N} +
      \gamma (T_\rL - p_{-N}^2(t)) \right) \dd t + \sqrt{2\gamma T_\rL} p_{-N}(t) \, \dd W_\rL(t), \\
    \dd J^e_{N, N+1}(t) & = -\left(\tau_\rR p_N + \gamma (T_\rR - p_{N}^2(t))\right) \dd t
    - \sqrt{2\gamma T_\rR} p_{N}(t) \, \dd W_\rR(t).
\end{aligned}
\]

\paragraph{The stationary state.}
For $T_\rR = T_\rL = \beta^{-1}$ and $\tau_\rL = \tau_\rR = \tau$, there is a unique stationary probability measure:
\begin{equation}
  \label{eq:eq_measure}
\dd \nu^N_{\beta, \bar p} = \prod_{i=-N}^N \frac{\rme^{-\beta e_i +\beta \bar p p_i}}{Z_{\beta,\bar p}} \, \dd p_i \, \dd r_i,
\end{equation}
with $\bar p = \tau\gamma^{-1}$. This corresponds to an \emph{equilibrium} situation, for which $\left\langle j^a_{i-1,i}(t)\right\rangle_{N,\beta, \bar p} = 0$ with $a \in \{p,e\}$. The symbol $\left\langle \cdot \right\rangle_{N,\beta, \bar p}$ denotes the expectation with respect to~$\nu^N_{\beta, \bar p}$.

If $T_\rR \neq T_\rL$ or $\tau_\rL \neq \tau_\rR$, the stationary probability measure cannot be computed explicitly. In fact even the existence of an invariant probability measure is an open problem for chains of lengths greater than~4 (see~\cite{CEP15,CE16,CP17}).
In what follows, we assume the existence and uniqueness of the stationary state and
the expectation with respect to the stationary probability measure is denoted by
$\left\langle \cdot\right\rangle_{N,\ss}$.

Some identities are immediate consequences
of the conservation laws~\eqref{eq:1a}--\eqref{eq:1b}: for any $-N \leq i \leq N-1$,
\[
\begin{split}
\tau_\rL - \gamma \left \langle p_{-N}\right\rangle_{N,\ss} &= -\left\langle V'(r_{i+1})\right\rangle_{N,\ss} = -\tau_\rR + \gamma \left\langle p_{N}\right\rangle_{N,\ss}, \\
\tau_\rL \left\langle p_{-N}\right\rangle_{N,\ss} + \gamma \left( T_\rL - \left\langle p_{-N}(t)^2 \right\rangle_{N,\ss} \right) &= -\left\langle p_i V'(r_{i+1})\right\rangle_{N,\ss} = - \tau_\rR  \left\langle p_{N}\right\rangle_{N,\ss}-\gamma \left(T_\rR - \left\langle p_{N}(t)^2 \right\rangle_{N,\ss}\right).
\end{split}
\]
We expect the average currents to have a well defined \emph{thermodynamic} limit,
\emph{i.e.} that there exist some quantities $J^a\left(T_\rL,T_\rR,\tau_\rL,\tau_\rR\right) \in \mathbb{R}$
such that, for any $i \in \mathbb{Z}$, 
\begin{equation}
\label{eq:27}
\lim_{N\to\infty} N 
\left\langle j^a_{i,i+1}\right\rangle_{N,\ss} = J^a\left(T_\rL,T_\rR,\tau_\rL,\tau_\rR\right),
\qquad
a \in \{p,e\},
\end{equation}
which implies the diffusive behavior of the conserved quantities. 
We also expect that there exist two functions~$p_\ss, e_\ss : [-1,1] \to \mathbb{R}$ such that 
\begin{equation}
\label{eq:stsprof}
\forall x \in [-1,1], \qquad \lim_{N\to\infty} \left\langle p_{[Nx]} \right\rangle_{N,\ss} =  p_{\ss}(x) , \qquad
\lim_{N\to\infty} \left\langle e_{[Nx]} \right\rangle_{N,\ss} =  e_{\ss}(x). 
\end{equation}
The functions $p_{\ss}(x)$ and $e_{\ss}(x)$
are the stationary profiles of momentum and energy, respectively, and are solutions of a stationary diffusive system.
As we expect the system to be locally at equilibrium, 
there is equivalently a stationary temperature profile
$T_\ss : [-1,1] \to \mathbb{R}_+$ defined as
\begin{equation}
\label{eq:temp}
\lim_{N\to\infty} \left\langle p_{[Nx]}^2 \right\rangle_{N,\ss} -\left\langle p_{[Nx]} \right\rangle^2_{N,\ss} = T_{\ss}(x).
\end{equation}
Moreover, the following boundary conditions should be satisfied:
\begin{equation}
\label{eq:bc}
p_{\ss}(-1) = p_\rL, \quad p_{\ss}(1) = p_\rR, \quad T_{\ss}(-1) = T_\rL, \quad T_{\ss}(1) = T_\rR.
\end{equation}

\paragraph{Entropy production.}
We derive inequalities that allow to determine the signs of the average currents. We first construct a reference Gibbs measure~$\widetilde \nu^N$, and rewrite the stationary probability measure as $f_\ss^N \widetilde \nu^N$ in order to define a relative entropy based on~$f_\ss^N$. The reference measure is an inhomogeneous Gibbs measure associated with profiles $\mathfrak{b},\mathfrak{bp} : [-1,1] \to \mathbb{R}$ of local values of the thermodynamic parameters conjugate to energy and momentum: 
\begin{equation}
\label{eq:gibbs-grad}
\dd\widetilde \nu^N = \prod_{i=-N}^N Z_{\mathfrak{b}(i/N), \mathfrak{bp}(i/N)}^{-1} \exp\left(-\mathfrak{b}\left(\frac iN\right) e_i + \mathfrak{bp}\left(\frac iN\right) p_i\right) \, \dd p_i \, \dd r_i.
\end{equation}
We choose linear interpolation profiles: denoting by $\beta_\rR = T_\rR^{-1}$ and $\beta_\rL = T_\rL^{-1}$,
\[
\mathfrak{b}\left(x\right) = \frac{\beta_\rL+\beta_\rR}{2} + x \frac{\beta_\rR-\beta_\rL}{2},
\qquad
\mathfrak{b p}\left(x\right) = \frac{\beta_\rL p_\rL + \beta_\rR p_\rR}{2} + x \frac{\beta_\rR p_\rR-\beta_\rL p_\rL}{2}.
\]
A simple computation shows that
\begin{equation}
\label{eq:11}
0 = \left\langle L_{N,T_\rL,T_\rR,p_\rL,p_\rR} \left(\log f_{\ss}^N\right)\right\rangle_{N, \ss} = \int A_N  f_{\ss}^N \, \dd \widetilde \nu^N + \gamma \int f_{\ss}^N \left[ \left(\S_{\rL} + \S_{\rR}\right) \log f_{\ss}^N \right] \dd \widetilde \nu^N .
\end{equation}
Observe that, by integration by parts, 
\[
\int A_N f_{\ss}^N \,  \dd \widetilde \nu^N = (\beta_\rR-\beta_\rL) \left\langle \frac1{2N} \sum_{i=-N+1}^N j^e_{i-1,i} \right\rangle_{N,\ss} - \frac{\beta_\rR \tau_\rR-\beta_\rL\tau_\rL}{\gamma} \left\langle \frac1{2N} \sum_{i=-N+1}^N j^p_{i-1,i} \right\rangle_{N,\ss}.
\]
Denoting by $O^*$ the adjoint of a closed operator~$O$ on $L^2(\widetilde \nu^N)$,
we have that $\S_{\rL} = -T_\rL \partial_{p_{-N}}^* \partial_{p_{-N}}$ and
$\S_{R} = -T_\rR \partial_{p_{N}}^* \partial_{p_{N}}$, therefore these operators are 
symmetric on $L^2(\widetilde \nu^N)$ and
\[
  \gamma \int  f_{\ss}^N \left(\S_{\rL} + \S_{\rR}\right) \log f_{\ss}^N \, \dd \widetilde \nu^N
=  -\gamma T_\rL\int \frac{\left(\partial_{p_{-N}}f_{\ss}^N\right)^2}{f_{\ss}^N} \, \dd \widetilde \nu^N - \gamma T_\rR \int \frac{\left(\partial_{p_N}f_{\ss}^N \right)^2}{f^N_{\ss}} \, \dd \widetilde\nu^N := -\widetilde\sigma_N,
\]
where we introduce the entropy production~$\widetilde\sigma_N$.
From~\eqref{eq:11} and the fact that the currents are uniform in space, we obtain that the stationary state satisfies the following entropy production inequality
\begin{equation}
  \label{eq:22}
  \forall i = -N+1,\dots,N, \qquad \widetilde\sigma_N = 
  \left( \beta_\rR-\beta_\rL\right) \left\langle j^e_{i-1,i} \right\rangle_{N,\ss} - \left(\beta_\rR p_\rR-\beta_\rL p_\rL\right) \left\langle j^p_{i-1,i} \right\rangle_{N,\ss} \ge 0,
\end{equation}
\emph{i.e.}
\[
\left(T_\rL - T_\rR \right) \left\langle j^e_{i_-1,i}\right\rangle_{N,\ss} - \gamma^{-1}\left(T_\rL \tau_\rR - T_\rR \tau_\rL \right) \left\langle j^p_{i-1,i}\right\rangle_{N,\ss} \ge 0.
\]
As discussed in Section~\ref{ssec:ssproperties}, the above inequality provides
a lower bound on the energy current, and imposes constraints on the region of parameter space
where uphill energy diffusion can be observed. 

\subsection{Linear response and the Onsager matrix}
\label{ssec:onsager-matrix}

In this section we define the transport coefficients associated with small variations in the average currents arising from perturbations of the equilibrium state. These transport coefficients are given by Green--Kubo formulas. We first consider the linear response of the average currents for a system of finite size started close to the stationary state, and then perform a formal large space-time limit. 

\paragraph{Linear response of average currents.}
We consider a system initialized at time $t=0$ with the inhomogeneous measure~$\widetilde\nu^N$ defined in~\eqref{eq:gibbs-grad}, and introduce some small variations
\[
\eps^e = \frac{\beta_\rR-\beta_\rL}{2}, \qquad \eps^p = \frac{\beta_\rR p_\rR-\beta_\rL p_\rL}{2}.
\]
Linear response theory suggests that the average currents at time~$t \geq 0$ are linearly related at dominant order to the variations~$\eps^e,\eps^p$:
\begin{equation}
\label{eq:linres}
\begin{split}
N \langle j_{0,1}^p(t) \rangle_{\widetilde\nu^N} &= -K^{p,p}_{N}(t) \eps^p  + K^{p,e}_{N}(t) \eps^e  + \rmo(|\eps^e|,|\eps^p|), \\
N \langle j_{0,1}^e(t) \rangle_{\widetilde\nu^N} &= -K^{e,p}_{N}(t) \eps^p + K^{e,e}_{N}(t) \eps^e
+ \rmo(|\eps^e|,|\eps^p|),
\end{split}
\end{equation}
where the expectation is taken with respect to initial conditions distributed according to~$\widetilde\nu^N$ and for all realizations of the nonequilibrium dynamics~\eqref{eq:dyn_bulk}--\eqref{eq:dyn_boundary}.

We first take the limit $ N \to +\infty$ and assume that $N \langle j_{0,1}^a(t) \rangle_{\widetilde \nu^N} \xrightarrow[N\to+\infty]{} \mathcal{J}^a(t)$ (with $a \in \{e,p\}$);
then the limit $t\to\infty$ and assume that $\mathcal{J}^a(t) \xrightarrow[t\to+\infty]{} J^a$.
We assume in addition that the response coefficients $K^{a,b}_{N}(t)$, $a,b \in \{e,p\}$, also have limits $K^{a,b}$ when $N\to +\infty$ and $t \to +\infty$,
and that the error $\rmo(|\eps^e|,|\eps^p|)$ remains uniform in~$t$ and~$N$. Then,
\[
\begin{split}
J^p =  -K^{p,p} \eps^p  + K^{p,e} \eps^e + \rmo(|\eps^e|,|\eps^p|), \\
J^e = -K^{e,p} \eps^p + K^{e,e} \eps^e + \rmo(|\eps^e|,|\eps^p|).
\end{split}
\]
The matrix whose coefficients are $K^{a,b}$, is the so-called Onsager matrix.
We next identify concretely the formal expressions of the limit response coefficients~$K^{a,b}$. 

A straightforward expansion at first order in~$\eps^e,\eps^p$ of the probability measure~$\widetilde\nu^N$ and of the evolution semigroup $\rme^{t L_{N,T_\rL,T_\rR,p_\rL,p_\rR}}$ as
\[
\begin{aligned}
& \dd\widetilde\nu^N = \left(1 - \sum_{i=-N}^N \frac{i}{N} \left[\eps^e \left(e_i-\left\langle e_i \right\rangle_{N,\beta,\bar{p}}\right) - \eps^p \left( p_i-\left\langle p_i \right\rangle_{N,\beta,\bar{p}} \right)\right] \right)\dd\nu^N_{\beta, \bar p} + \mathrm{O}\left(|\eps^e|^2,|\eps^p|^2\right), \\
& \rme^{t L_{N,T_\rL,T_\rR,p_\rL,p_\rR}} = \rme^{t L_{N,\beta^{-1},\beta^{-1},\bar{p},\bar{p}}} + \mathrm{O}(|\eps^e|,|\eps^p|),
\end{aligned}
\]
allows to express the response coefficients~$K^{a,b}_{N}(t)$ at dominant order as
\[ 
\begin{split}
  K^{p,p}_{N}(t) = -\sum_{i=-N}^N i \left\langle j_{0,1}^p(t) \left(p_i(0)- \left\langle p_i \right\rangle_{N,\beta, \bar p}\right) \right\rangle_{N,\beta, \bar p},
  & \quad
  K^{p,e}_{N}(t) = - \sum_{i=-N}^N i \left\langle j_{0,1}^p(t) \left(e_i(0)- \left\langle e_i \right\rangle_{N,\beta, \bar p}\right) \right\rangle_{N,\beta, \bar p}, \\
  K^{e,p}_{N}(t) =  -\sum_{i=-N}^N i \left\langle j_{0,1}^e(t) \left(p_i(0)- \left\langle p_i \right\rangle_{N,\beta, \bar p}\right) \right\rangle_{N,\beta, \bar p},
  & \quad
  K^{e,e}_{N}(t) = - \sum_{i=-N}^N i \left\langle j_{0,1}^e(t) \left(e_i(0)- \left\langle e_i \right\rangle_{N,\beta, \bar p}\right) \right\rangle_{N,\beta, \bar p},
\end{split}
\]
where the expectation is taken with respect to the equilibrium probability measure~\eqref{eq:eq_measure} with the same temperatures $\beta^{-1} = (T_\rL + T_\rR)/2$ and the same forcings $\gamma \bar p = (\beta_\rL\tau_\rL + \beta_\rR\tau_\rR)/(2\beta)$ at the boundaries, and for all realizations of the associated equilibrium dynamics~\eqref{eq:dyn_bulk}--\eqref{eq:dyn_boundary}. 

Thanks to the symmetries of the equilibrium dynamics with respect to time reversal and rotations, we can rewrite the transport coefficients~$K^{a,b}_N(t)$ in a form more suitable for taking the limits $N \to +\infty$ and $t \to +\infty$. Denoting by $\mathcal{R}$ the momentum reversal operator, namely
\[
(\mathcal{R}\phi)(r_{-N+1},\dots,r_N,p_{-N},\dots,p_N) = \phi(r_{-N+1},\dots,r_N,-p_{-N},\dots,-p_N),
\]
it holds (with the short-hand notation $L_{N,\beta^{-1},\overline{p}} = L_{N,\beta^{-1},\beta^{-1},\overline{p},\overline{p}}$)
\[
\left\langle \left(L_{N,\beta^{-1},\overline{p}}\phi\right) \varphi \right\rangle_{N,\beta,\overline{p}} = \left\langle \mathcal{R}\phi\left(L_{N,\beta^{-1},-\overline{p}}\mathcal{R}\varphi\right)  \right\rangle_{N,\beta,-\overline{p}}.
\]
Therefore, by time reversal symmetry,
\begin{equation}
\label{eq:time_symmetry}
\left\langle \phi(t) \varphi \right\rangle_{N,\beta,\overline{p}} := \left\langle \left(\rme^{t L_{N,\beta^{-1},\overline{p}}}\phi\right) \varphi \right\rangle_{N,\beta,\overline{p}} = \left\langle \mathcal{R}\phi\left(\rme^{t L_{N,\beta^{-1},-\overline{p}}}\mathcal{R}\varphi\right) \right\rangle_{N,\beta,-\overline{p}} = \left\langle \mathcal{R}\phi \, (\mathcal{R}\varphi)(t) \right\rangle_{N,\beta,-\overline{p}},
\end{equation}
where the last average is taken
for all realizations of the equilibrium dynamics~\eqref{eq:dyn_bulk}--\eqref{eq:dyn_boundary}
with the same temperature~$\beta^{-1}$ and $-\tau = -\gamma \bar{p}$ (instead of~$\tau$)
at both boundaries.
As for the rotational symmetry of the equilibrium dynamics, it holds
\begin{equation}
  \label{eq:rotational_symmetry}
  \left\langle \rme^{tL_{N,\beta^{-1},\overline{p}}} \theta_{\overline{p}}\phi, \theta_{\overline{p}}\varphi \right\rangle_{N,\beta,\overline{p}} = \left\langle \rme^{tL_{N,\beta^{-1},0}} \phi, \varphi  \right\rangle_{N,\beta,0},
\end{equation}
where $(\theta_{\overline{p}}\phi)(r_{-N+1},\dots,r_N,p_{-N},\dots,p_N) = \phi(r_{-N+1},\dots,r_N,p_{-N}-\overline{p},\dots,p_N-\overline{p})$. The proof of this identity relies on the observation that $\theta_{-\overline{p}}\, L_{N,\beta^{-1},\overline{p}}\, \theta_{\overline{p}} = L_{N,\beta^{-1},0}$.

\paragraph{Expression of $K^{p,p}$.}
Using $\left\langle j_{0,1}^p(0) p_i(0)\right\rangle_{\beta, \bar p} = 0$
and the time symmetry property~\eqref{eq:time_symmetry} for $\phi(q,p)= j^p_{0,1}(q,p)$ and $\varphi(q,p)=p_i$, and then the rotational symmetry~\eqref{eq:rotational_symmetry}, we obtain
\begin{align*}
  K^{p,p}_{N}(t) &= -\sum_{i=-N}^N i \left\langle j_{0,1}^p(t) p_i(0)\right\rangle_{N,\beta, \bar p} = \sum_{i=-N}^N i \left\langle j_{0,1}^p(0) p_i(t)\right\rangle_{N,\beta, -\bar p} = \sum_{i=-N}^N i \left\langle j_{0,1}^p(0) (p_i(t) - p_i(0))\right\rangle_{N,\beta, -\bar p} \\
  & = \sum_{i=-N}^N i \left\langle j_{0,1}^p(0) (p_i(t) - p_i(0))\right\rangle_{N,\beta,0}.
\end{align*}
Now, in view of \eqref{eq:1a}--\eqref{eq:1b},
\begin{align*}
  K^{p,p}_{N}(t) &= \int_0^t \sum_{i=-N+1}^{N-1} i \left\langle j_{0,1}^p(0) (j^p_{i-1,i}(s)
                   - j^p_{i,i+1}(s))\right\rangle_{N,\beta,0} \dd s \\
& \qquad - N \left\langle \int_0^t j_{0,1}^p(0) (\dd J^p_{-N-1,-N}(s) - j^p_{-N,-N+1}(s) \, \dd s)\right\rangle_{N,\beta,0} \\
& \qquad + N \left\langle \int_0^t j_{0,1}^p(0) (j^p_{N-1,N}(s) \, \dd s - \dd J^p_{N,N+1}(s))\right\rangle_{N,\beta,0} \\
&= \int_0^t \sum_{i=-N}^{N-1}  \left\langle j_{0,1}^p(0) j^p_{i,i+1}(s)\right\rangle_{N,\beta,0} \dd s - B_{N}^{p,p}(t),
\end{align*}
where
\begin{equation*}
    \begin{split}
  B_{N}^{p,p}(t) =  N \left\langle  j_{0,1}^p(0) \left( J^p_{-N-1,-N}(t)
      +  J^p_{N,N+1}(t) \right) \right\rangle_{N,\beta,0} = \gamma N \int_0^t \left\langle  V'(r_0(0))
    \left(p_{-N}(s) - p_{N}(s)\right) \right\rangle_{N,\beta,0} \, \dd s.
\end{split}
\end{equation*}
By the locality of the dynamics, we expect that 
$N\left\langle  V'(r_0(0)) p_{\pm N}(s) \right\rangle_{N,\beta,0}\to 0$ as $N\to\infty$,
which implies
\begin{equation}
\label{eq:8}
 \lim_{N\to\infty} B_{N}^{p,p}(t) = 0. 
\end{equation}
Finally, assuming that the large space-time limits are well defined, we obtain the following Green-Kubo formula for~$K^{p,p}(\beta, \bar p)$:
\begin{equation}
\label{eq:9}
\begin{split}
K^{p,p}(\beta, \bar p)  & := \lim_{t\to \infty}  \lim_{N\to\infty} K^{p,p}_{N}(t)
= \int_0^\infty \sum_{i\in \mathbb Z}  \left\langle j_{0,1}^p(0) j^p_{i,i+1}(s)\right\rangle_{\beta, 0} \dd s
= K^{p,p}(\beta, 0) := K^{p,p}(\beta),
\end{split}
\end{equation}
where $\langle \cdot\rangle_{\beta, \bar p}$ denotes the expectation for the infinite dynamics at equilibrium with parameters~$\beta,\bar p$. Note that we can formally deduce that $K^{p,p}$ is nonnegative from the space-time invariance of the infinite volume equilibrium dynamics:
\[
K^{p,p}(\beta)  = \lim_{t,N \to +\infty}\left\langle \frac{1}{4Nt} \left( \int_0^t \sum_{i=-N}^N j^p_{i,i+1}(s)\right)^2 \right\rangle_{\beta, 0} \geq 0.
\]

\paragraph{Expression of the other response coefficients.}
The other response coefficients are defined by Green--Kubo formulas similar to~\eqref{eq:9}. Applying first the rotational symmetry~\eqref{eq:rotational_symmetry}, and then the time reversal symmetry~\eqref{eq:time_symmetry},
\begin{equation}
  \label{eq:1}
  \begin{aligned}
    K^{e,p}_{N}(t)
    & = -\sum_{i=-N}^N i \left\langle j_{0,1}^e(t) p_i(0) \right\rangle_{N,\beta, \bar p}
    = -\sum_{i=-N}^N i \left\langle j_{0,1}^e(t) p_i(0) \right\rangle_{N,\beta,0} - \bar{p} \sum_{i=-N}^N i \left\langle j_{0,1}^p(t) p_i(0) \right\rangle_{N,\beta,0} \\
    & = -\sum_{i=-N}^N i \left\langle j_{0,1}^e(0) p_i(t) \right\rangle_{N,\beta,0} + \bar{p} K^{p,p}_N(t) = -\sum_{i=-N}^N i \left\langle j_{0,1}^e(0) \left(p_i(t) - p_i(0)\right) \right\rangle_{N,\beta,0} + \bar{p} K^{p,p}_N(t) \\
    & = -\int_0^t \sum_{i=-N}^N \left\langle j_{0,1}^e(0) j_{i,i+1}^p(s) \right\rangle_{N,\beta,0} \dd s +B^{e,p}_N(t) + \bar{p} K^{p,p}_N(t),
\end{aligned}
\end{equation}
where we assume as in~\eqref{eq:8} that $B^{e,p}_N(t) \to 0$ as $N\to \infty$. In the limit $t\to +\infty$, this gives the definition
\begin{equation}
  \label{eq:6}
  K^{e,p}(\beta, \bar p)  = K^{e,p}(\beta,0) + \bar{p} K^{p,p}(\beta),
  \qquad 
  K^{e,p}(\beta,0) = -\int_0^\infty\sum_{i\in \mathbb Z}\left\langle j^e_{0,1}(0)j^p_{i,i+1}(s)\right\rangle_{\beta,0} \, \dd s.
\end{equation}
Similarly,
\begin{equation}
  \label{eq:7}
  \begin{aligned}
    K^{p,e}_{N}(t) & =  -\sum_{i=-N}^N i \left\langle j_{0,1}^p(t) e_i(0) \right\rangle_{N,\beta, \bar p}
    = -\sum_{i=-N}^N i \left\langle j_{0,1}^p(t) e_i(0) \right\rangle_{N,\beta,0} + \bar{p} K^{p,p}_N(t) \\
    & = -\sum_{i=-N}^N i \left\langle j_{0,1}^p(0) e_i(t) \right\rangle_{N,\beta,0} + \bar{p} K^{p,p}_N(t)
    = -\sum_{i=-N}^N i \left\langle j_{0,1}^p(0) \left(e_i(t) - e_i(0)\right) \right\rangle_{N,\beta,0} + \bar{p} K^{p,p}_N(t) \\
    & = -\int_0^t \sum_{i=-N}^N \left\langle j_{0,1}^p(0) j_{i,i+1}^e(s) \right\rangle_{N,\beta,0} \dd s
  +B^{p,e}_N(t) + \bar{p} K^{p,p}_N(t),
  \end{aligned}
\end{equation}
where we also assume that $B^{p,e}_N(t) \to 0$ as $N\to \infty$, which leads to defining
\begin{equation}
  \label{eq:6-1}
  K^{p,e}(\beta, \bar p) = K^{p,e}(\beta,0) + \bar{p} K^{p,p}(\beta),
  \qquad
  K^{p,e}(\beta,0) = -\int_0^\infty\sum_{i\in \mathbb Z}\left\langle j^p_{0,1}(0)j^e_{i,i+1}(t)\right\rangle_{\beta,0} \, \dd t.
\end{equation}
Since $K^{p,e}(\beta,0) = K^{e,p}(\beta,0)$ by the time reversal symmetry, the following Onsager reciprocal relation holds: 
\begin{equation}
  K^{p,e}(\beta, \bar p) = K^{e,p}(\beta,\bar p).
\end{equation}
In an analogous way, we obtain that
\begin{equation}
\label{eq:9bis}
K^{e,e}(\beta, \bar p) = \int_0^\infty \sum_{i\in \mathbb Z}\left\langle j^e_{0,1}(t)j^e_{i,i+1}(0)\right\rangle_{\beta, \bar p} \, \dd t \geq 0.
\end{equation}
In view of the rotational symmetry, it holds
\[
\begin{split}
  K^{e,e}(\beta,  \bar p) & = \int_0^\infty\sum_{i\in \mathbb Z}\left\langle (p_0(0) + \bar p)(p_i(t) + \bar p) j^p_{0,1}(0) j^p_{i,i+1}(t)\right\rangle_{\beta, 0} \, \dd t \\
  & = \int_0^\infty\sum_{i\in \mathbb Z}\left\langle p_0(0) p_i(t) j^p_{0,1}(0) j^p_{i,i+1}(t)\right\rangle_{\beta, 0} \, \dd t + \bar p^2 \int_0^\infty\sum_{i\in \mathbb Z}\left\langle j^p_{0,1}(0) j^p_{i,i+1}(t)\right\rangle_{\beta, 0} \, \dd t \\
  & = K^{e,e}(\beta, 0) + \bar p^2  K^{p,p}(\beta).
\end{split}
\]

\paragraph{Expressions of the Onsager coefficients for even potentials.}
We conclude this section by giving a simpler expression of the Onsager coefficients for potentials~$V$ which are even functions of~$r$. For such potentials, the equilibrium distribution on the path space for $\bar p = 0$ is invariant (i.e. symmetric) with respect to the sign flip of all coordinates $\{ (r_i(t),p_i(t) \}_{i \in \mathbb{Z},t\in[0,T]} \to \{ (-r_i(t),-p_i(t)) \}_{i \in \mathbb{Z},t\in[0,T]}$. This implies that, for all $t \ge 0$,
\[
  \left\langle p_0(0) j^p_{0,1}(0)j^p_{i,i+1}(t) \right\rangle_{\beta,0} = 0,
\]
since the function $p_0(0) j^p_{0,1}(0)j^p_{i,i+1}(t)$ is antisymmetric with respect to the sign flip of all coordinates. In this case,
\[
K^{e,p}(\beta,0) = K^{p,e}(\beta,0) = 0.
\]
In conclusion, for even potentials~$V$ the Onsager matrix reads
\begin{equation}
  \label{eq:Onsager_matrix}
\begin{pmatrix} K^{p,p}(\beta, \bar p) & K^{p,e}(\beta, \bar p) \\ K^{e,p}(\beta, \bar p) & K^{e,e}(\beta, \bar p) \end{pmatrix} = K^{p,p}(\beta) \begin{pmatrix} 1 & \bar p \\ \bar p & \bar p^2 \end{pmatrix} + K^{e,e}(\beta,0) \begin{pmatrix} 0 & 0 \\ 0 & 1 \end{pmatrix}.
\end{equation}

\subsection{Equations for the stationary profiles}
\label{ssec:stationary-profiles}

The linear response framework described in the previous section gives the first order term of the currents
for a perturbation of the equilibrium created by gradients of temperature and momentum.
Since we have defined $J^a$ by \eqref{eq:27}, the currents can be rewritten as
\begin{equation}
  \label{eq:27b}
\lim_{N\to\infty} N
\left\langle j^a_{[Nx],[Nx]+1}\right\rangle_{N,\ss} = J^a,
\qquad
a \in \{p,e\}, \quad x\in [-1,1]. 
\end{equation}
On the other hand, around a \emph{macroscopic} point $x\in [-1,1]$, the equilibrium is perturbed by the \emph{local} gradients of the temperatures and momentum. 
In view of~\eqref{eq:linres}, for any $x\in [-1,1]$ the equations for the stationary profiles
defined in~\eqref{eq:stsprof} should read
\begin{equation}\label{eq:ss1}
\begin{split}
J^p &= -K^{p,p}(\beta_{\ss}(x)) \partial_x\big( \beta_{\ss}(x) p_{\ss}(x) \big) + K^{p,e}(\beta_{\ss}(x),p_{\ss}(x)) \partial_x\beta_{\ss}(x), \\
J^e &= -K^{e,p}(\beta_{\ss}(x),p_{\ss}(x)) \partial_x\big(\beta_{\ss}(x)p_{\ss}(x)\big) + K^{e,e}(\beta_{\ss}(x),p_{\ss}(x)) \partial_x\beta_{\ss}(x). 
\end{split}
\end{equation}
A rigorous derivation of~\eqref{eq:ss1} would require a difficult hydrodynamic
limit in the stationary state, involving a proof of local equilibrium and its first order correction.

Thanks to the properties of the Onsager matrix elements~\eqref{eq:Onsager_matrix}, we can simplify the expressions of the currents as
  \begin{equation}
    J^p = -K^{p,p}(\beta_\ss) \partial_x\left(\beta_\ss p_\ss\right)
    + K^{p,p}(\beta_\ss) p_\ss \partial_x\beta_\ss = -\beta_\ss K^{p,p}(\beta_\ss) \partial_x p_\ss,
    \label{eq:ssp}
\end{equation}
and
  \begin{equation}
  \begin{split}
    J^e & = K^{p,p}(\beta_\ss) p_\ss \partial_x\left(\beta_\ss p_\ss\right)
    + \left(K^{e,e}(\beta_\ss) - p_\ss^2 K^{p,p}(\beta_\ss)\right) \partial_x\beta_\ss \\ 
    & = K^{p,p}(\beta_\ss)  \beta_\ss \partial_x \left(\frac{p_\ss^2}2\right)
    + K^{e,e}(\beta_\ss)  \partial_x\beta_\ss.
  \end{split}\label{eq:sse}
\end{equation}
In order to express the currents in terms of the temperature profile $T_\ss= \beta_\ss^{-1}$, we define the momentum diffusivity~$D^p$ and the the thermal conductivity~$\kappa$ as
\begin{equation}
  \label{eq:diffusive_coeff}
D^p(T) = \frac 1T K^{p,p}\left(\frac1T\right) \ge 0, \qquad \kappa(T) = \frac 1{T^2} K^{e,e}\left(\frac1T\right) \ge 0 .
\end{equation}
With this change of variable, the equations satisfied by the stationary profiles are
\begin{equation}
\label{eq:19}
\begin{split}
J^p &= - D^{p}(T_{\ss})\partial_x p_{\ss}, \\
J^e &= - D^{p}(T_{\ss}) \partial_x\left( \frac{p_{\ss}^2}2\right) - \kappa(T_{\ss}) \partial_xT_{\ss},
\end{split}
\end{equation}
where the values of $J^p, J^e$ are determined by the boundary conditions~\eqref{eq:bc}.
Note that the energy current is the sum of the \emph{heat current}
\begin{equation}
  \label{eq:JQ}
  J^Q(x) = - \kappa(T_{\ss}(x)) \partial_x T_{\ss}(x),
\end{equation}
and the \emph{mechanical energy current}
\begin{equation}
  \label{eq:Jm}
  J^m(x) = -D^p(T_{\ss}) \partial_x\left(\frac{p_{\ss}^2}2\right) = p_{\ss}(x) J^p,
\end{equation}
which can be of opposite signs and cause the occurrence of \emph{uphill diffusion}~\cite{Krishna15} of energy. This phenomenon arises when the mechanical current dominates the heat current, that is when $J^e$ has the same sign as the gradient of temperature (see Section~\ref{ssec:ssproperties}). 

\subsubsection{First properties of the diffusive system}

We assume the existence and the uniqueness of the solutions of~\eqref{eq:19}, and provide simple explicit bounds. From the first equation in~\eqref{eq:19}, we remark  that $\partial_x p_\ss$ has a constant sign (opposite to the sign of~$J^p$), hence~$p_\ss$ is a monotonic function and the following maximum principle holds
\[
  \min\left(p_\rL,p_\rR\right) \le p_{\ss}(x) \le  \max\left(p_\rL,p_\rR\right). 
\]

In order to obtain bounds on the temperature profile, we introduce an antiderivative $\mathcal K(T)$ of $\kappa(T)$ (\emph{i.e.} a
strictly increasing function such that $\mathcal K'(T) =\kappa(T)$). The second equation in~\eqref{eq:19} can then be rewritten as
\begin{equation}
  \label{eq:12}
  \partial_x \left[ \mathcal{K}(T_\ss(x)) \right] = p_{\ss}(x) J^p - J^e,
\end{equation}
or, in integral form,
\[
  \mathcal{K}(T_\ss(x)) = \mathcal{K}(T_\rL) + J^p \int_{-1}^x p_{\ss}(y) \, \dd y - J^e(x+1)
  = \mathcal{K}(T_\rR) - J^p \int_{x}^1 p_{\ss}(y) \, \dd y + J^e(1-x).
\]
This implies that
\[
  \sup_{x\in [-1,1]} |\mathcal{K}(T_\ss(x))| \le \max\left( \left|\mathcal{K}(T_\rL)\right|, \left|\mathcal{K}(T_\rR)\right|\right)
  +\max\left( \left|p_\rR\right|,\left|p_\rL\right|\right) |J^p| +  |J^e|,
\]
hence we can derive bounds on $T_\ss$ by applying the inverse function~$\mathcal{K}^{-1}$ to both sides of the above inequality.
  
\subsubsection{Symmetry properties}
\label{sec:symm-prop-solut}

System~\eqref{eq:19} has various symmetry properties, which allow to restrict the range of momentum and temperature boundary values to be explored and to simplify some mathematical arguments for the qualitative study of the solutions of~\eqref{eq:19} (see Section~\ref{ssec:ssproperties}). These properties are:
\begin{enumerate}[label=(\alph*)]
\item \textbf{Symmetry by rotation:} for fixed temperatures $T_\rL,T_\rR$,
  the quantities $\partial_x p_{\ss}$, $J^p$ and~$T_\ss$
  depend only on the difference $\tau_\rR - \tau_\rL$.
  Indeed, if we change $(\tau_\rL,\tau_\rR)$ to $(\tau_\rL + \bar \tau,\tau_\rR + \bar \tau)$,
  \eqref{eq:19} implies the change of $p_{\ss}(x)$ to $p_{\ss}(x) +\bar p$
  with $\bar p = \bar\tau/\gamma$ and $J^e$ to $J^e + \bar p J^p$.
  The other quantities remaining unchanged.
  \item \textbf{Symmetry by inversion of external force:} When changing $(\tau_\rL,\tau_\rR)$ to
    $(-\tau_\rL,-\tau_\rR)$, the stationary profile of temperature~$T_\ss$ and the energy current $J^e$ do not change, while~$p_\ss$ and $J^p$ change sign.
  \item \textbf{Symmetry by boundary exchange (trivial):}
    When exchanging ~$(\tau_\rL,T_\rL)$ with $(\tau_\rR,T_\rR)$, the new profiles are
  $\widetilde T_\ss(x) = T_\ss(-x)$ and $\widetilde p_\ss(x) = p_\ss(-x)$, and the currents change signs.
\end{enumerate}
The above properties show that the quantity of interest is in fact $\Delta p = p_\rR - p_\rR$, not the particular values we assign to $p_\rR$ and $p_\rL$ individually. This is why we consider $\tau_\rL=0$ and $p_\rR = \overline{p} = \overline{\tau}/\gamma$ in most of our numerical experiments, without loss of generality.

The elementary symmetries we discussed imply other symmetries: for instance, an exchange of the boundary temperatures leads to modified profiles~$\widetilde{p}_\ss,\widetilde{T}_\ss$ which can be obtained from the reference profiles $p_\ss,T_\ss$ as $\widetilde p_\ss(x) = \overline{p} - p_\ss(-x)$, $\widetilde T_\ss(x) = T_\ss(-x)$, $\widetilde J^p =J^p$ and $\widetilde J^e = -J^e + \overline{p} J^p$; while exchanging the boundary forces implies $\widetilde p_\ss(x) = \overline{p} - p_\ss(x)$, $\widetilde T_\ss(x) = T_\ss(x)$, $\widetilde J^p =-J^p$ and $\widetilde J^e = J^e - \overline{p} J^p$.

\begin{remark}
  \label{sec:rmk_symmetry_TL=TR}
  In the case $T_\rR = T_\rL$, it follows directly from the above symmetries that the profile of temperature is always symmetric with respect to the vertical axis~$x=0$ (\emph{i.e.} $T_\ss(x) = T_\ss(-x)$), for any possible values of~$p_\rL$ and $p_\rR$.
\end{remark}

\subsection{Analytical properties of the stationary state}
\label{ssec:ssproperties}
In this section we analyse the macroscopic entropy production in the stationary state (Section~\ref{sec:macr-entr-prod}) and list some qualitative properties of the stationary states, such as the shape of the profiles, with the characterization of their extremal and inflection points, and their behavior in the low temperature limit (Section~\ref{sec:maximum-inflection-other}).

\subsubsection{Macroscopic entropy production}
\label{sec:macr-entr-prod}

In the limit for $N \to \infty$, the macroscopic entropy production $\Sigma$ converges to
\begin{align}
  \Sigma = \lim_{N\to\infty} N \widetilde\sigma_N & = \left(T_\rR^{-1} - T_\rL^{-1}\right) J^e - \gamma^{-1} \left(T_\rR^{-1}\tau_\rR - T_\rL^{-1}\tau_\rL\right) J^p \label{eq:Sigma_compute} \\ 
         & = T_\rR^{-1} J^Q(1) -  T_\rL^{-1} J^Q(-1), \notag
\end{align}
where~$\widetilde\sigma_N$ is given by~\eqref{eq:22}.
Note that $\Sigma$ depends only on $T_\rL$, $T_\rR$, $\partial_x T_\ss\big|_{x=-1}$ and $\partial_x T_\ss\big|_{x=1}$. Since
\[
\partial_x \left[\frac{J^Q(x)}{T_\ss(x)}\right] = \frac{\kappa(T_{\ss}(x))}{T_{\ss}^2(x)} \left(\partial_x T_{\ss}(x)\right)^2 + \frac{D^p(T_{\ss}(x)) }{T_{\ss}(x)} \left(\partial_x p_{\ss}(x)\right)^2,
\]
we obtain that
\begin{equation}
  \label{eq:31}
  \Sigma = \int_{-1}^{1} \left[\frac{\kappa(T_{\ss}(x))}{T_{\ss}^2(x)} \left(\partial_x T_{\ss}(x)\right)^2 + \frac{D^p(T_{\ss}(x)) }{T_{\ss}(x)} \left(\partial_x p_{\ss}(x)\right)^2 \right] \dd x.
\end{equation}
Note that $\Sigma$ is invariant under all the symmetries described in Section~\ref{sec:symm-prop-solut}.

The resulting macroscopic entropy production inequality implies
  \[
  \left( \beta_\rR-\beta_\rL\right) J^e \geq \left(\beta_\rR p_\rR- \beta_\rL p_\rL\right) J^p,
  \]
and since $V'$ is bounded, we obtain
  \[
    \left( \beta_\rR-\beta_\rL \right) J^e \geq -\left|\beta_\rR p_\rR- \beta_\rL p_\rL\right|\, \sup_{r \in \mathbb{T}} |V'|.
  \]
  Therefore, $J^e$ cannot be too large in absolute value when it has the same sign as $T_\rR-T_\rL$, \emph{i.e.} in the presence of uphill energy diffusion (recall that in the absence of mechanical forcing, $J^e$ has the same sign as $T_\rL-T_\rR$ with the usual convention, so the previous inequality is trivially satisfied). 

 \subsubsection{Stationary points and other qualitative properties of the stationary profiles}
\label{sec:maximum-inflection-other}

This section lists various analytical properties of the solutions to~\eqref{eq:19}, which will facilitate the interpretation of the numerical results presented in Section~\ref{sec:numerics} -- see in particular the profiles in Figures~\ref{fig:Fig02}, \ref{fig:Fig04}, \ref{fig:Fig08} and~\ref{fig:Fig14}. 

For sufficiently large values of the boundary momentum~$\bar p$, a global temperature maximum appears in the bulk of the system, and the profile becomes more and more peaked as $\bar p$ increases or $T_\rL, T_\rR$ decrease. At the same time, the momentum profiles steepen, with a fast transition arising in the vicinity of the temperature maximum. 
We present some analytical arguments to explain these facts, starting with certain properties of the temperature maximum and then discussing the shape of the temperature profile.

\paragraph{Maxima of temperature.}
We distinguish two situations: (i) the temperature maximum is at the boundaries, as in the case of systems subjected only to a thermal forcing; (ii) the temperature maximum is in the interior of the domain and higher than the boundary values, as a result of the coupled thermal and mechanical forcings. 

We start by showing that there are only maxima of temperature (and no local minima). Since the antiderivative~$\mathcal{K}$ of~$\kappa$ is increasing, the stationary points of~$T_\ss$ coincide with those of~$\mathcal{K}(T_\ss)$. By~\eqref{eq:12}, a stationary point~$x_{\rm stat}$ must satisfy
\[
  p_\ss(x_{\rm stat}) = \frac{J^e}{J^p}.
\]
Since $p_\ss$ is strictly monotonic when~$\Delta p \neq 0$, there exists at most one stationary point. A necessary and sufficient condition for the existence of this stationary point is
\begin{equation}
  \label{eq:15}
  \min\left(p_\rL,p_\rR\right) \le \frac{J^e}{J^p} \le \max\left(p_\rL,p_\rR\right).
\end{equation}
By derivation with respect to~$x$ of equation  \eqref{eq:12}, we obtain
\begin{equation}\label{eq:19b}
  J^p \partial_x p_{\ss}(x) = \kappa(T_{\ss}(x)) \partial_{x}^2 T_{\ss}(x)
  + \kappa'(T_{\ss}(x)) \left(\partial_{x}T_{\ss}(x)\right)^2.
\end{equation}
Therefore, a stationary point of $T_{\ss}$ is always a maximum, since $\partial_x^2 T_\ss <0$ whenever $\partial_x T_\ss=0$ ($J^p  \partial_x p_{\ss}(x) < 0$ by~\eqref{eq:19}).

We denote by~$x_{_{T_\ss^{\rm max}}}$ the point where the maximum of temperature is attained.  When $T_\rR = T_\rL$, the maximum is obtained at $x_{_{T_\ss^{\rm max}}}=0$, by the symmetry properties of the profiles (see Remark~\ref{sec:rmk_symmetry_TL=TR}). In fact, in this case,
\[
  \int_{-1}^1 p_\ss(x) \, \dd x = \frac{2J^e}{J^p}
\]
and condition \eqref{eq:15} is satisfied for all values of $p_\rL, p_\rR$. When the temperature maximum is in the interior of the domain (namely $x_{_{T_\ss^{\rm max}}} \in (-1,1)$) and~$T_\rL \neq T_\rR$, its position is on the side of the highest boundary temperature, for any value of $p_\rL,p_\rR$. 

\begin{figure}
    \begin{center}
      \includegraphics[width=0.49\textwidth]{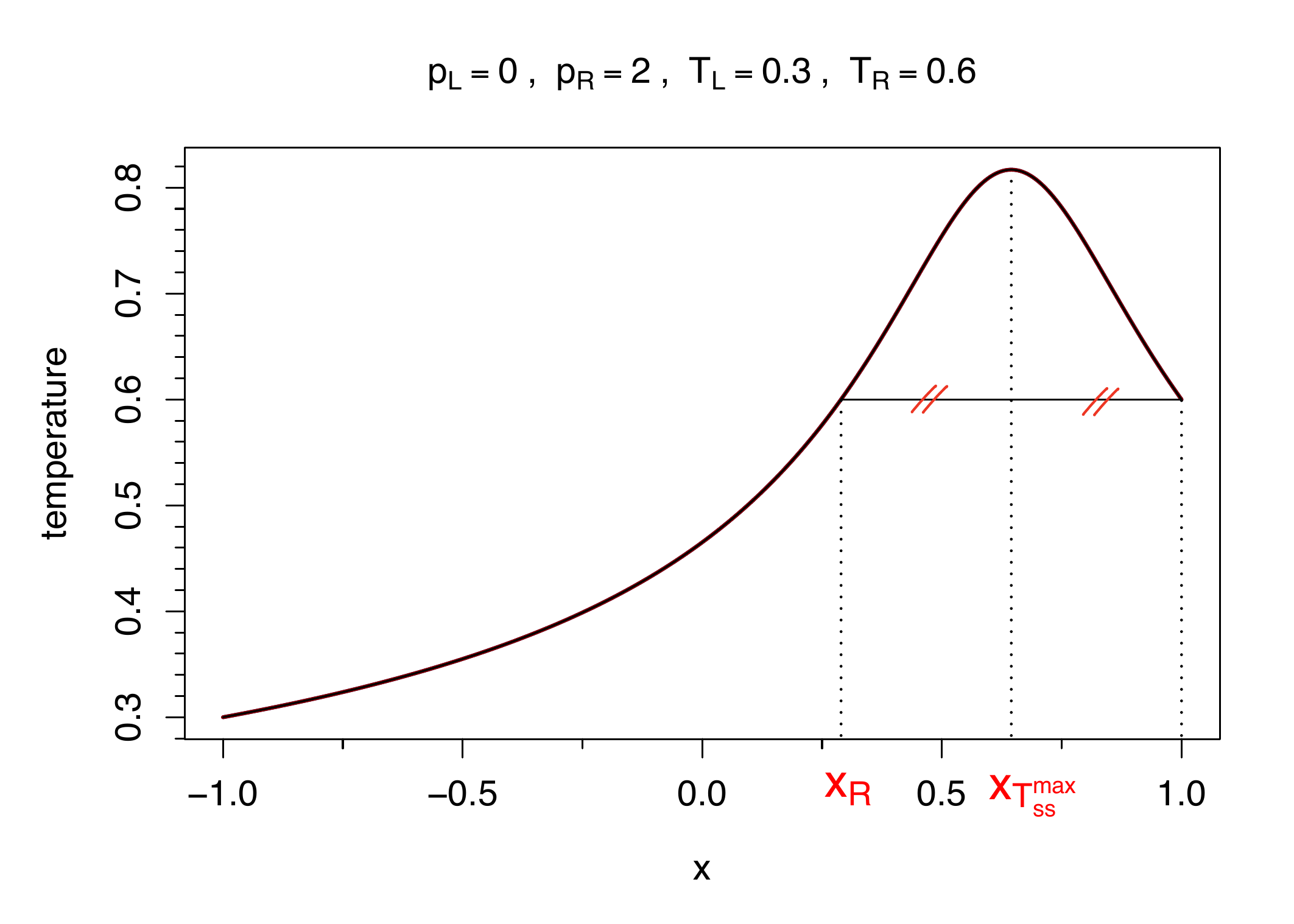}
      \includegraphics[width=0.49\textwidth]{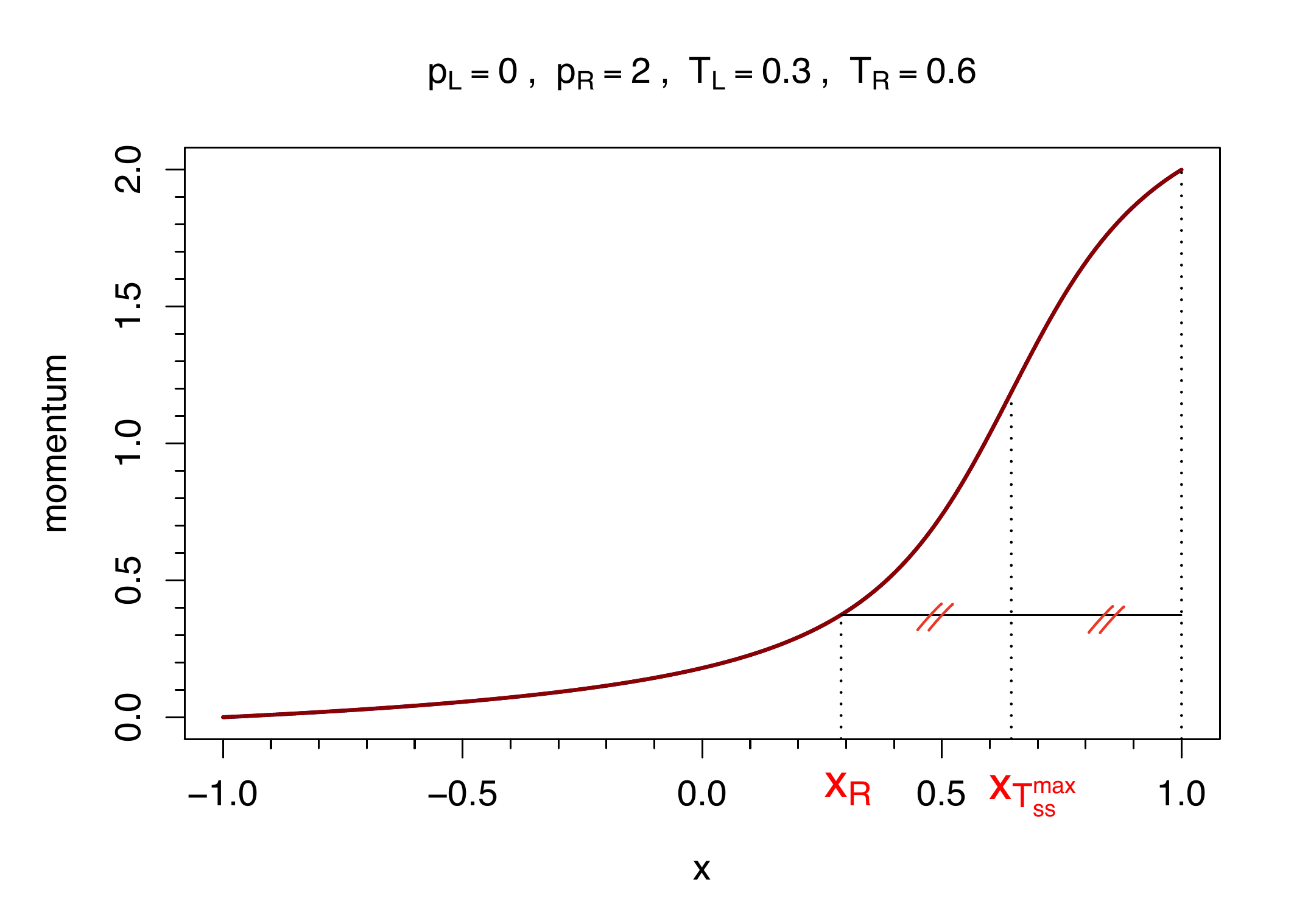}
      \caption[$T_\ss$ symmetry]{\small Local symmetry properties of~$T_\ss$ and $p_\ss$. The profiles are obtained by numerical integration of~\eqref{eq:19}, see details in Appendix~\ref{App:SSnum}.
      }		
      \label{fig:Fig000} 
    \end{center}
  \end{figure}

Moreover, when $x_{_{T_\ss^{\rm max}}} \in (-1,1)$, $T_\ss$ is symmetric with respect to $x=x_{_{T_\ss^{\rm max}}}$, \emph{i.e.} $T_\ss(x_{_{T_\ss^{\rm max}}}+ y) = T_\ss(x_{_{T_\ss^{\rm max}}}- y)$ for any $y \in \mathbb R_+$.
To prove the latter statements, consider for instance the case $T_\rL < T_\rR$. We refer to Figure~\ref{fig:Fig000}, where we show profiles of temperature and a momentum obtained by numerical integration of~\eqref{eq:19} (see Appendix~\ref{App:SSnum}). We denote by~$x_{\rR}$ the only element $-1 < x_\rR < x_{_{T_\ss^{\rm max}}}$ such that $T_\ss(x_\rR) = T_\rR$, and we set $\widetilde{p}_\rL = p_\ss(x_\rR)$.  
  Then, the profiles~$\widetilde{p}_\ss$ and~$\widetilde{T}_\ss$
  defined on the interval~$[x_\rR,1]$ satisfy~\eqref{eq:19}, with equal temperatures at the boundaries~$\widetilde{T}_\ss(x_\rR) = \widetilde{T}_\ss(1) = T_\rR$, and boundary momenta~$\widetilde{p}_\ss(x_\rR) = \widetilde{p}_\rL$ and~$\widetilde{p}_\ss(1) = p_\rR$. By the temperature profile symmetry discussed in Remark~\ref{sec:rmk_symmetry_TL=TR}, the temperature maximum is in the middle of the interval~$[x_\rR,1]$, \emph{i.e.} $x_{_{T_\ss^{\rm max}}} = (1+x_\rR)/2$, so that $0<x_{_{T_\ss^{\rm max}}}<1$. In the case $T_\rL> T_\rR$, by a similar reasoning we obtain that, if a maximum exists in~$(-1,1)$, its abscissa is such that~$-1 < x_{_{T_\ss^{\rm max}}}< 0$. 

\paragraph{Profile shapes.}
When boundary momenta increase at fixed $T_\rL,T_\rR$, we observe numerically that the temperature profiles become more peaked, while the momenta profiles steepen (see Figures~\ref{fig:Fig02} and~\ref{fig:Fig04}). Our simulations also show that, when boundary temperatures increase at fixed $p_\rL,p_\rR$, the temperature profiles become more spiky and their maximum raises and shifts towards the hotter boundary (see Figures~\ref{fig:Fig08} and~\ref{fig:Fig14}). 

Since the temperature maximum is observed in a region where the profile curvature is negative, we turn our attention to inflection points where the curvature vanishes:
\begin{enumerate}[label=(\arabic*)]
\item By derivation with respect to~$x$ of the first equation in~\eqref{eq:19}, we obtain
\[
  \partial_{x}^2 p_{\ss} (x) = J^p \frac{(D^p)'(T_{\ss}(x))}{D^p(T_{\ss}(x))^2} \partial_x T_{\ss}(x).
\]
This implies that $p_{\ss}$ has an inflection point at the temperature profile maximum.
\item By~\eqref{eq:19b}, the inflection point $x_{_{T_\ss^{\rm flex}}}$ of $T_{\ss}$ must satisfy
\begin{equation*}
  \kappa'\left(T_{\ss}\left(x_{_{T_\ss^{\rm flex}}}\right)\right) \left(\partial_{x}T_{\ss}\left(x_{_{T_\ss^{\rm flex}}}\right)\right)^2 = - D^p\left(T_{\ss}\left(x_{_{T_\ss^{\rm flex}}}\right)\right)  \left(\partial_x p_{\ss}\left(x_{_{T_\ss^{\rm flex}}}\right)\right)^2 \leq 0.
\end{equation*}
It follows that such inflection points can only exist in regions where $\kappa(T)$ is a strictly decreasing function of $T$. We observe numerically that $\kappa(T)$ is a decreasing function of $T$ for our model, at least in the considered range of temperatures (see Section~\ref{sec:description} and Figure~\ref{fig:Fig01}).
\end{enumerate}

\paragraph{Low temperature limit.}
We consider $p_\rL=0$, $p_\rR=\bar p$ and define~$T_{\rm crit}$ as the critical temperature at which the thermal conductivity diverges. We prove that, if the temperatures at the boundaries decrease to~$T_{\rm crit}$, 
the temperature profile spikes at its maximum but remains bounded, which seems consistent with the numerical results obtained in~\cite{ILLP14}.
We assume that~$\kappa,D^p$
are non-increasing functions of the temperature and that, for $T > T_{\rm crit}$ given,
\begin{equation}
\lim_{\quad \bar{T} \to T_{\rm crit}^+} \int_{\bar{T}}^{T} \kappa(\theta) \, \dd \theta = +\infty.
\label{eq:kappadiv}
\end{equation}
The latter equation implies that $\kappa(\bar T) \to +\infty$ as $\bar T\to T_{\rm crit}^+$, but also a stronger property, ensuring that the divergence of the thermal conductivity is sufficiently strong. This guarantees that the temperatures remain above~$T_{\rm crit}$ (see Remark~\ref{rmk:on_kappadiv} below for further comments).

The microscopic simulation results presented presented in Section~\ref{sec:description} prompt the divergence of the thermal conductivity at low temperatures, in agreement with some previous studies (see \emph{e.g.}~\cite{GS00}), and legitimize the assumption that both~$\kappa$ and $D^p$ are non-increasing. 
The authors of~\cite{GS00} suggest that $T_{\rm crit}>0$ for rotors, but their simulations are based on quite short chains, thus it cannot be take for granted that the thermodynamic limit at low temperatures is attained. We believe that $T_{\rm crit} = 0$ for rotors, but we emphasize that the value of~$T_{\rm crit}$ is irrelevant for the argument we present here.  

We consider for simplicity the symmetric situation $T_\rR = T_\rL = \bar T$ and $p_\rR = - p_\rL = \bar p >0$. Then $J^p <0$ and the temperature maximum is at~$x_{_{T_\ss^{\rm max}}}= 0$, where the momentum is~$p_\ss(0) = 0$. This implies that $J^e= p_\ss(0) J^p = 0$. First note that, by setting~$J^e = 0$ in the second equation in~\eqref{eq:19} and integrating over~$[0,1]$, the maximum value of the stationary temperature can be written as
\[
  T_\ss(0) = \bar T + \int_0^1 \frac{\left|J^p\right| p_\ss(y)}{\kappa(T_\ss(y))} \dd y.
\]
We next extend the solution of~\eqref{eq:19} to the interval $[-\lambda, \lambda]$ with $\lambda >1$:
\[
  -p_\ss(-\lambda) = p_\ss(\lambda) = \bar p + \int_1^\lambda \partial_x p_\ss(x) \, \dd x =
  \bar p + \int_1^\lambda \frac{\left|J^p\right|}{D^p(T_\ss(x))} \, \dd x.
\]
Since $D^p$ is non-increasing by assumption, and $x \mapsto T_\ss(x)$ is non-increasing for $x\ge 0$, we have that
\[
\bar p +  \frac{\left|J^p\right| (\lambda-1)}{D^p(T_\ss(\lambda))} \le p_\ss(\lambda) \le  \bar p +  \frac{\left|J^p\right| (\lambda-1)}{D^p(\bar T)}.
\]
From~\eqref{eq:12}, we also obtain that
\begin{equation}
  \label{eq:mathcalK_lambda}
  \int_{T_\ss(\lambda)}^{\bar{T}} \kappa = |J^p|\int_1^\lambda p_\ss.
\end{equation}
Since $p_\ss(x) \ge \bar p$ for $x\in [1,\lambda]$, the integral on the right-hand side goes to infinity as $\lambda \to +\infty$,
which implies by~\eqref{eq:kappadiv} that $T_\ss(\lambda)$ remains above $T_{\rm crit}$, and 
\[
  \lim_{\lambda\to \infty} T_\ss(\lambda) = T_{\rm crit}.
\]
We now rescale the extended profile back to the interval $[-1,1]$ by defining
\[
  p_\ss^\lambda (y) =  p_\ss (\lambda y) , \qquad
  T_\ss^\lambda (y) =  T_\ss (\lambda y), 
\]
which satisfy the equations
\begin{equation}
  \label{eq:30}
  \begin{aligned}
    \lambda J^p & = -D^p\left(T_\ss^\lambda (y)\right) \partial_y p_\ss^\lambda (y), \\
    0 &= -D^p\left(T_\ss^\lambda (y)\right) \partial_y \left(\frac{p_\ss^\lambda (y)^2}2\right)
    - \kappa\left(T_\ss^\lambda (y)\right) \partial_y T_\ss^\lambda (y),
  \end{aligned}
\end{equation}
with boundary conditions $p_\ss^\lambda (1) = p_\ss(\lambda) = - p_\ss^\lambda (-1) \geq \bar{p}$
and $T_\ss^\lambda (1) = T_\ss(\lambda) = T_\ss^\lambda (-1) \le \bar{T}$. They are therefore solutions of the stationary equations~\eqref{eq:19} with the latter boundary conditions and such that $T_\ss^\lambda (\pm 1) \to T_{\rm crit}$ as $\lambda \to +\infty$. In fact, this extension-rescaling procedure is a way to keep the temperature maximum~$T_\ss(0)$ fixed when decreasing the boundary temperature towards~$T_{\rm crit}$ while simultaneously increasing the value of the boundary momenta.

Thanks to the above result, showing that~$T_\ss(0)$ is a non decreasing function of~$p_\ss(1) = -p_\ss(-1)$ is sufficient to prove our initial claim, \emph{i.e.} that the temperature maximum is bounded when the the boundary temperature tends to the critical temperature with boundary momenta fixed. In fact, by
integrating the second equation in~\eqref{eq:30}, one obtains
\begin{equation}
  \label{eq:T_increasing_p}
p_\ss(1)^2 = 2 \int_{\bar T}^{T_\ss(0)} \frac{\kappa(\theta)}{D^p(\theta)} d\theta,
\end{equation}
which implies that~$T_\ss(0)$ is an increasing function of~$p_\ss(1)^2$.

\begin{remark}
  \label{rmk:on_kappadiv}
  It is not sufficient to assume that $\kappa(T) \to +\infty$
  as $T\to T_{\rm crit}^+$ in order to ensure that the temperature always remains above $T_{\rm crit}$ in the extension process we describe above.
  Indeed, assume that $D^p(\theta) = 1$ and $\kappa(\theta) = \kappa_0 \theta^{-\alpha}$ for some
  $\alpha \in (0,1)$ and fix $T_{\rm crit} = 0$.
  Since $p_\ss(x) \ge \bar{p}$ for $x\in [1,\lambda]$, it follows from~\eqref{eq:mathcalK_lambda} that
  \[
    \frac{\kappa_0}{1-\alpha} \left(\bar{T}^{1-\alpha}-T_\ss(\lambda)^{1-\alpha}\right) =
    |J^p| \int_1^\lambda p_\ss (y) \dd y \ge  |J^p|\bar{p}(\lambda-1).
  \]
  In particular, there exists $\lambda^* > 1$ such that $T_\ss(\lambda^*) = 0$, and the solution cannot be continued after this value. This issue does not arise when $\alpha \geq 1$, a condition guaranteed by~\eqref{eq:kappadiv}. 
\end{remark}

To conclude, since 
\[
\partial_x p_\ss^\lambda (x) = \lambda \partial_x p_\ss(\lambda x),
\qquad
\partial_x T_\ss^\lambda (x) = \lambda \partial_x T_\ss(\lambda x),
\qquad \partial_x^2 T_\ss^\lambda (x) = \lambda^2 \partial_x^2 T_\ss(\lambda x),
\]
it follows that the momentum profiles steepen as $\lambda \to \infty$, while the temperature profiles peak around their maximum value. Indeed, the first equation in~\eqref{eq:30} leads to
\[
  \partial_y p_\ss^\lambda (0) = \frac{\lambda \left|J^p\right|}{D^p(T_\ss(0))}
  \mathop{\longrightarrow}_{\lambda \to \infty} +\infty,
\]
while, by differentiating the second equation in~\eqref{eq:30} with respect to~$y$ and taking into account that the temperature derivative vanishes at~$y=0$,
\[
  \partial^2_y T_\ss^\lambda (0) = - \frac{\lambda \left|J^p\right| \partial_y p_\ss^\lambda (0)}{\kappa(T_\ss(0))}
  \mathop{\longrightarrow}_{\lambda \to \infty} -\infty,
\]
which shows the spiky behavior of $T_\ss^\lambda $ around $0$ as $\lambda \to \infty$.

\subsection{Qualitative discussion of phenomena induced by thermo-mechanical forcings}
\label{sec:qualitative_phenomena}

Note first that, by integrating the first equation in~\eqref{eq:19} with respect to $x$, we obtain, with the notation $\Delta p = p_\rR-p_\rL$ introduced in Section~\ref{sec:symm-prop-solut},
\begin{equation}
	J^p = - \left(\int_{-1}^1 \dfrac{1}{D^p(T_\ss)}\right)^{-1} \Delta p,
	\label{eq:24}
\end{equation}
so that, for all possible choices of boundary conditions for temperature and momentum, $J^p$ always has the same sign as~$-\Delta p$. Next, by integrating the second equation in~\eqref{eq:19} with respect to $x$, we obtain, with the notation $\Delta T = T_\rR-T_\rL$, 
\begin{equation}
	J^e = -\dfrac{1}{\ds\int_{-1}^1 \dfrac{1}{\kappa(T_\ss)}} \left[\Delta T + \dfrac{\ds\int_{-1}^1 \dfrac{p_\ss}{\kappa(T_\ss)} }{\ds\int_{-1}^1 \dfrac{1}{D^p(T_\ss)}} \, \Delta p \right].
	\label{eq:25}
\end{equation}

In what follows we discuss some interesting phenomena concerning the currents~$J^e,J^p$, which arise from the thermo-mechanical forcing. We first give some alternative expressions of these currents (Section~\ref{sec:alt_curr}), before discussing uphill energy diffusion (Section~\ref{sec:uphill-energy-diff}) and negative energy conductivity (Section~\ref{sec:neg-cond-diff}). 

\subsubsection{Alternative expressions of the currents}
\label{sec:alt_curr}

Since the currents $J^e,J^p$ are constant in space, they can be expressed using the derivatives of the temperature fields at the boundaries as
\begin{equation}
\begin{split}
J^e &= -\kappa(T_\rL) \partial_x T_\ss\big|_{x=-1} + p_\rL J^p 
= -\kappa(T_\rR) \partial_x T_\ss\big|_{x=1} + p_\rR J^p,
\end{split}
\label{eq:condizia}
\end{equation}
from which, considering~\eqref{eq:24}, we deduce that 
\begin{equation}
\partial_x T_\ss \big|_{x=-1} = \frac{\kappa(T_\rR)}{\kappa(T_\rL)} \partial_x T_\ss\big|_{x=1}
+ \left( \kappa(T_\rL)\int_{-1}^{1} \frac{1}{D^p(T_\ss)} \right)^{-1} \Delta p^2.
\label{eq:reldxTLdxTR}
\end{equation}
Note also that, when a maximum of $T_\ss$ is present in the interval $[-1,1]$, the following upper or lower bounds on the temperature derivatives at the boundaries can be obtained from~\eqref{eq:reldxTLdxTR}:
\[
  \begin{aligned}
0 & \le  \partial_x T_\ss\big|_{x=-1} \le \left( \kappa(T_\rL)\int_{-1}^{1} \frac{1}{D^p(T_\ss)} \right)^{-1} \Delta p^2, \\
0 & \ge  \partial_x T_\ss\big|_{x=1} \ge - \left( \kappa(T_\rR)\int_{-1}^{1} \frac{1}{D^p(T_\ss)} \right)^{-1} \Delta p^2.
\end{aligned}
\]

\subsubsection{Uphill energy diffusion}
\label{sec:uphill-energy-diff}

When $\Delta p \neq 0$, $\Delta T \neq 0$ and $p_\ss(x)\Delta p\Delta T < 0$ for all $x\in [-1,1]$, that is when the two terms in~\eqref{eq:25} have opposite signs, there may exist values of $(p_\rL,p_\rR)$ for which an uphill diffusion~\cite{Krishna15} of energy occurs. Indeed,
when there is no momentum gradient at the boundaries, the energy current $J^e$ equals the thermal current~$J^Q$, which has the same sign as~$-\Delta T= T_\rL-T_\rR$. On the other hand, as soon as $\Delta p = p_\rR-p_\rL \ne 0$, a mechanical current $J^m(x)$ appears, which has the same sign as~$-p_\ss(x)\Delta p$. When $J^e=J^m(x) + J^Q(x)$ is such that
\begin{equation}
  (T_\rR-T_\rL) J^e > 0,
  \label{eq:UHdef}
\end{equation}
the intensity of the mechanical forcing dominates that of thermal forcing and the resulting energy flow goes up the temperature gradient imposed at the boundary, towards the hottest thermostat, hence the uphill energy diffusion occurs (see Section~\ref{ssec:uphill} for a numerical illustration of this phenomenon). 
	
\subsubsection{Negative energy conductivity}
\label{sec:neg-cond-diff}

In systems satisfying Fourier's law and subjected only to a thermal forcing, the energy current~$J^e$ is proportional to~$-\Delta T$, so it has a positive linear response with respect to variations of~$-\Delta T$. In~\cite{ILOS11}, we first observed by numerical simulation of the microscopic dynamics described in Section~\ref{ssec:microdyn}, that for fixed values of~$p_\rR$ and $p_\rL$ the energy current $J^e$ may exhibit a \emph{negative} linear response with respect to variations of~$-\Delta T$, namely that $J^e$ may increase (decrease) when $\Delta T$ is increased (decreased)
\[
J^e \uparrow (\downarrow) \quad {\text as}\quad \Delta T \uparrow (\downarrow).
\]
Since the usual definition of thermal conductivity is given by the first order linear response to $-\Delta T$, we can talk about \emph{negative energy conductivity}.
  This should not be confused with the thermal conductivity coefficient $\kappa(T)$
  that is always positive: the local heat current $J^Q(x)$ always has the opposite sign
  of the local gradient of the temperature profile. In this sense, Fourier law is always satisfied locally.

Thanks to the macroscopic description~\eqref{eq:19}, we can understand this phenomenon as arising from the nonlinearity
of the transport coefficients (negative energy conductivity is not present in models where~$\kappa,D^p$ are constant, as in~\cite{kos19,kos20}) and stemming from the interplay of the two conserved quantities and the interaction between mechanical and thermal currents. However, we are only able to verify ex-post whether or not we are in a regime of parameters leading to negative energy conductivity, for example by interpreting temperature profiles in the light of~\eqref{eq:condizia}. No necessary, not to mention sufficient, conditions clearly appear from the qualitative analysis of~\eqref{eq:19} or from the numerical results discussed in Section~\ref{ssec:negTC}.
Yet, we believe that the presence of a stationary point in the temperature profile, which is necessarily a maximum, plays a central role in the emergence of negative energy conductivity, as it causes $\Delta T$ and $\partial_x T_\ss (x)$ to be of opposite signs (see {\it e.g.} the region $x \in (x_{_{T_\ss^{\rm max}}},1]$ in Figure~\ref{fig:Fig000}).

\section{Numerical investigation of the stationary state}
\label{sec:numerics}

We start by describing the numerical procedures used to compute the diffusion coefficients and the stationary profiles (Section~\ref{sec:description}), then we present the results showing the presence of uphill energy diffusion (Section~\ref{ssec:uphill}) and negative energy conductivity (Section~\ref{ssec:negTC}), for some regimes of parameters. 

\subsection{Description of the numerical method}
\label{sec:description}

\begin{figure}
  \begin{center}
    \includegraphics[height=0.4\textheight]{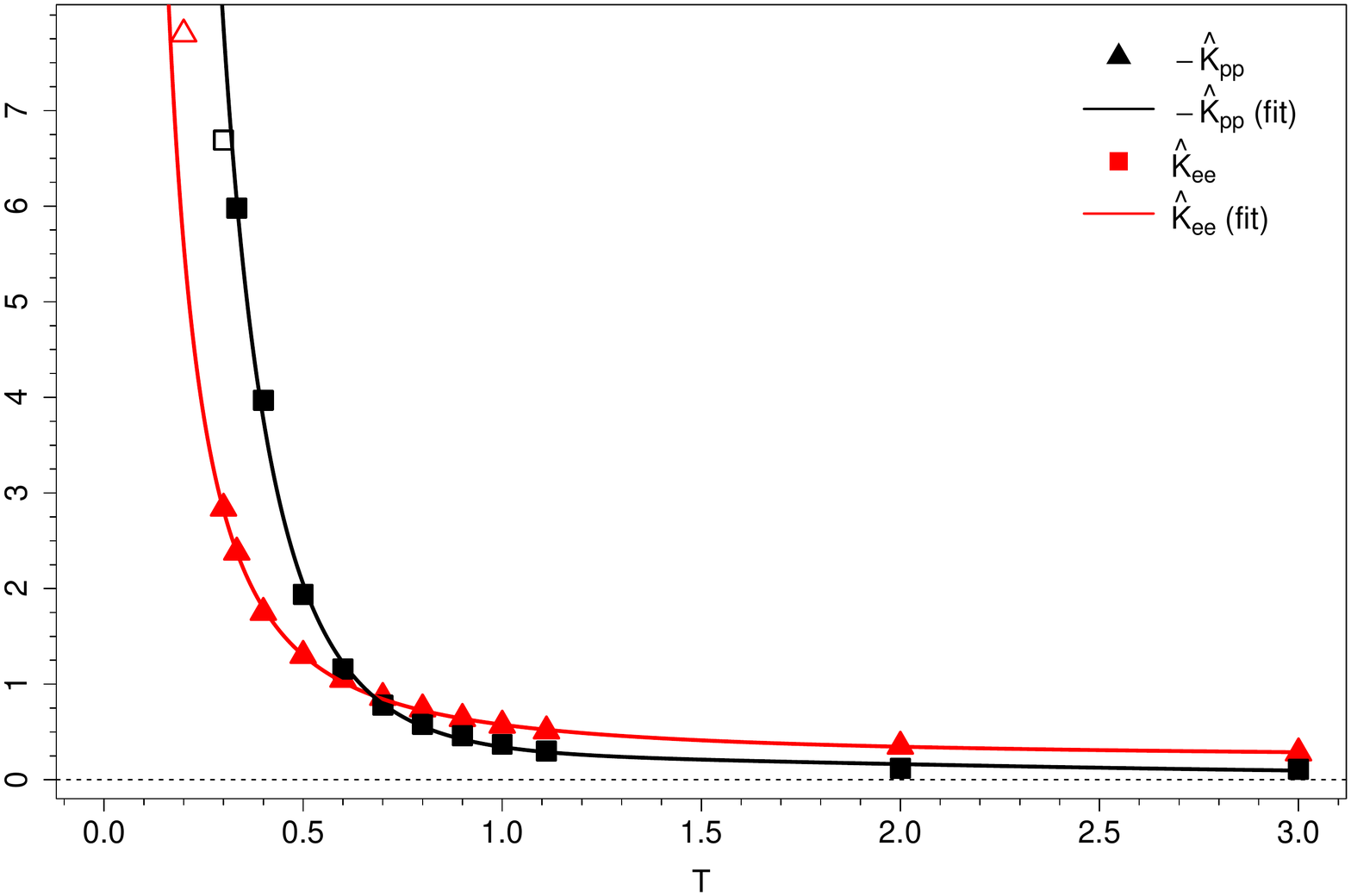}
    \caption[$\hat K^{p,p}$ and $\hat K^{e,e}$ data]{\small Estimated Onsager coefficients~$\hat K^{p,p}$ and~$\hat K^{e,e}$ and their fitting curves as functions of the temperature $T$. The empty points are those which are numerically less reliable and have thus not been considered in the fitting procedure.
    }		
    \label{fig:Fig01} 
  \end{center}
\end{figure}

\paragraph{Computation of the diffusion coefficients.}
In order to solve the diffusion system~\eqref{eq:19} numerically, we first need to estimate the diffusion coefficients~\eqref{eq:diffusive_coeff}, which amounts to estimating the Onsager coefficients~$K^{p,p}$ and~$K^{e,e}$ for various values of the temperature. This is done by numerically simulating the microscopic dynamics, following the procedure detailed in Appendix~\ref{App:GKnum}. The data points for the estimated coefficients~$\hat K^{p,p}$ and~$\hat K^{e,e}$ are reported in Figure~\ref{fig:Fig01}, together with their numerical fits.  The points corresponding to low temperature values ($T<0.3$) are not considered in the fitting procedure because their numerical estimation is not sufficiently accurate due to too long correlation times.

\begin{remark}
  The thermal conductivity, which is normal (\emph{i.e.}~it has a finite thermodynamic limit) for this system, has been estimated in~\cite{SG01} to be: (a\/)~$\kappa \approx 29.5$ for $T=0.3$ and $\kappa \approx 0.612$ for $T=1$, using nonequilibrium systems in the linear response regime; (\/b\/)~$\kappa \approx 28.5$ for $T=0.3$ and $\kappa \approx 0.55$ when $T=1$ using a Green--Kubo approach with Langevin dynamics for systems of size~$2N=4000$. The estimates we obtain by a Green--Kubo approach with Hamiltonian dynamics, are of the same order of magnitude, since $\kappa \approx 31.55$ for $T=0.3$ and $\kappa \approx 0.57$ for $T=1$.
\end{remark}

\paragraph{Numerical resolution of the diffusion system.}
Once the diffusion coefficients have been estimated via~\eqref{eq:fitGKcoeff}, we are able to numerically integrate the nonlinear system of stationary equations~\eqref{eq:19} using a finite difference discretization, together with a fixed-point algorithm to address the nonlinearity. The details are given in Appendix~\ref{App:SSnum}. 

\begin{figure}
  \begin{center}
    \includegraphics[width=0.49\textwidth]{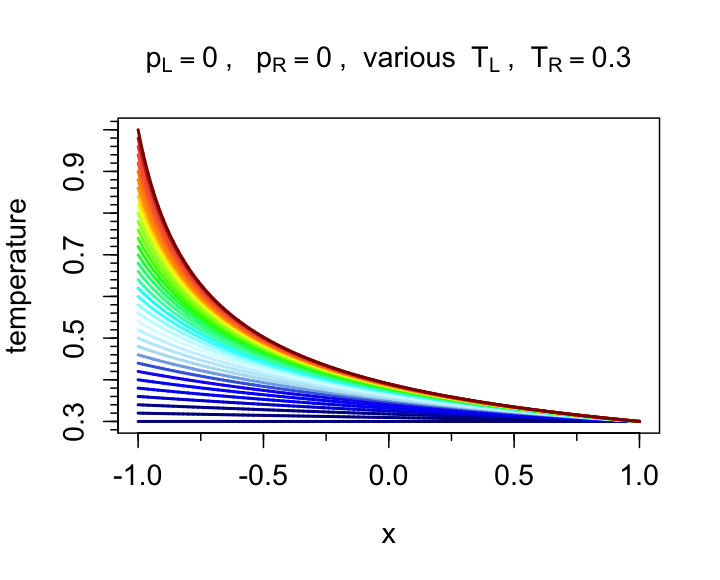}
    \includegraphics[width=0.49\textwidth]{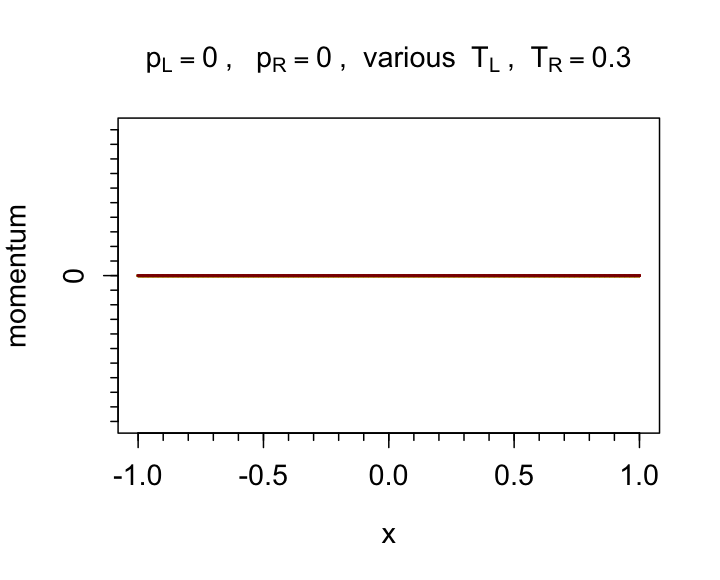}\\ 
    \includegraphics[width=0.49\textwidth]{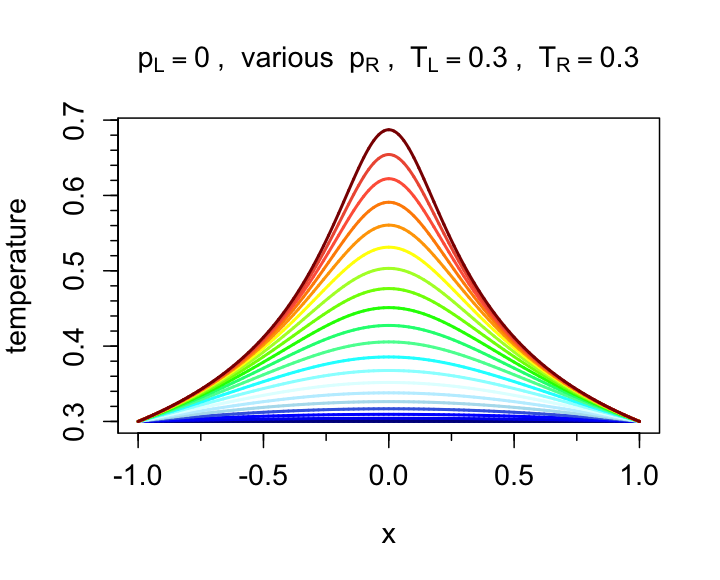}
    \includegraphics[width=0.49\textwidth]{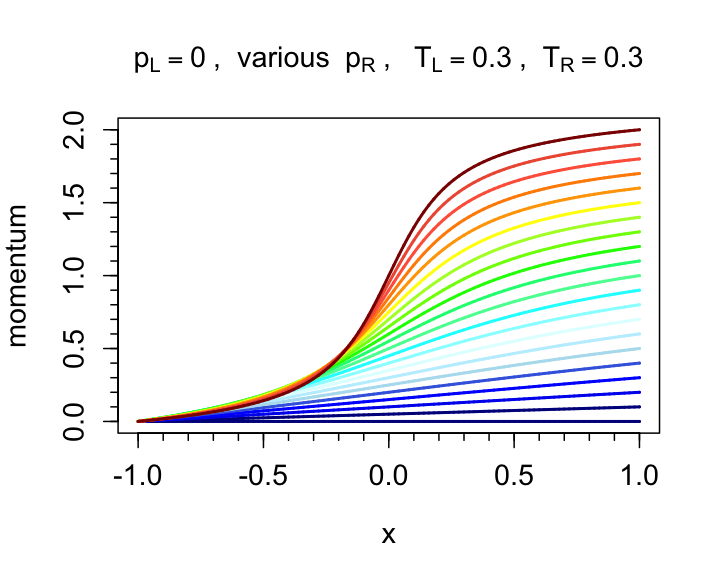}\\
    \caption[profiles with one gradient at the time]{\small Profiles of temperatures (left column) and momentum (right column). On the top row, we show the results for $p_\rL=p_\rR=0$ (no mechanical forcing), $T_\rR=0.3$  and values of $T_\rL$ varying from 0.3 (blue profiles) to 1 (dark-red profiles) in steps of~0.02. On the bottom row, we show the results  
      for $T_\rL = T_\rR = 0.3$ (no thermal forcing), $p_\rL=0$ and values of $p_\rR$ varying from 0 (blue profiles) to 2 (dark-red profiles) in steps of~0.1. Note that it suffices that only $\Delta p \neq 0$ to have a non-zero thermal current, arising solely from the dissipation of mechanical energy.} 		
    \label{fig:Fig02} 
  \end{center}
\end{figure} 
We show some profiles of temperature (left) and momentum (right) in Figure~\ref{fig:Fig02}. The plots on the top row correspond to the case $\Delta p=0$ and $\Delta T \leq 0$, in which the temperature profiles are solutions of the equation $J^e = - \kappa(T(x)) \partial_x T_\ss(x)$ and $p_\ss(x) = 0$. Since~$\kappa>0$ by definition and $J^e\geq 0$, the temperature profiles are nonincreasing functions of~$x$. Moreover, from~\eqref{eq:19b} considering~$J^p=0$ and $\Delta T\neq 0$, we obtain that~$\partial_x^2 T_\ss > 0$ when~$\kappa'(T_\ss)<0$, therefore the temperature profiles have a convex shape.

The plots on the bottom row in Figure~\ref{fig:Fig02} correspond to the case $\Delta T =0$ and $\Delta p \geq 0$. In this case, the temperature profile is symmetric with respect to the $y$-axis and presents a global maximum at $x=0$, as expected from the symmetry properties discussed in Section~\ref{ssec:stationary-profiles} (see Remark~\ref{sec:rmk_symmetry_TL=TR}). 
It is easy to see from~\eqref{eq:19} that the function $p_\ss$ is approximately affine for small values of $\Delta p$, and has variations of order~$\Delta p$, while $T_\ss$ is a concave parabola of order $(\Delta p)^2$. Nonlinearity increases in both profiles as $\Delta p$ is increased. When $\Delta p$ is sufficiently large, two inflection points appear in the temperature profile, so its curvature becomes negative in some regions, as discussed in Section~\ref{sec:maximum-inflection-other} (see in particular~\eqref{eq:19b}).

\subsection{Uphill energy diffusion}
\label{ssec:uphill}

Uphill energy diffusion~\eqref{eq:UHdef} may be observed in the system when $\Delta T \neq 0$, $\Delta p \neq 0$ and $p_\ss(x)\Delta p \Delta T <0$, as discussed in Section~\ref{sec:uphill-energy-diff}.
This corresponds to energy flowing from the coldest to the hottest thermostat.
The main cause of uphill diffusion is that the mechanical current eventually prevails on the heat current. For $\Delta p\neq 0$ large enough at given $\Delta T \neq 0$, this effect is combined with that caused by the presence of a maximum in the temperature profile, which contributes in returning some heat to the hottest thermostat --~by which we mean that~$J^Q(x)$ has the same sign as~$\Delta T$ in some regions. Of course Fourier's law remains valid since~$J^Q(x)$ remains proportional to~$-\partial_x T_\ss$.

\begin{figure}
  \begin{center}
    \includegraphics[height=0.25\textheight]{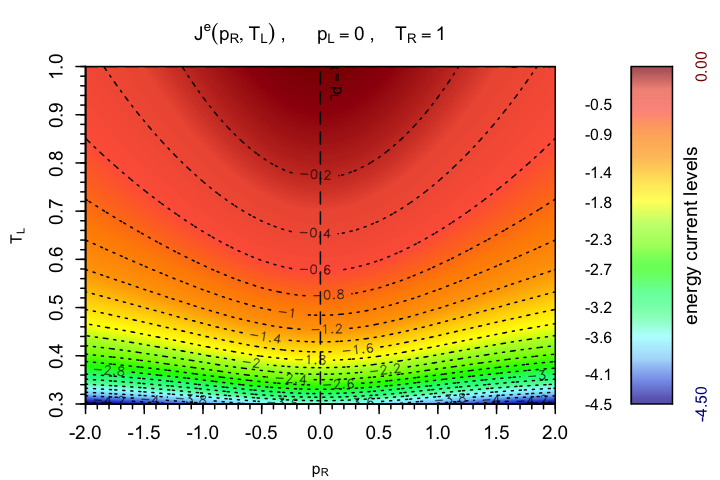}
    \includegraphics[height=0.25\textheight]{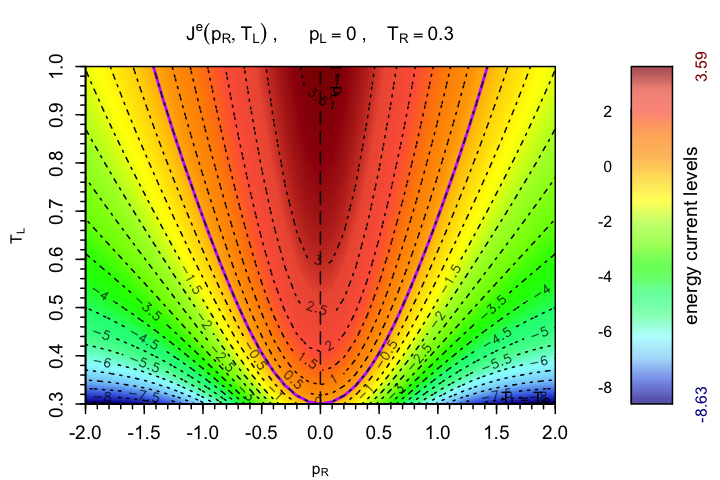}
    \caption[$J^e(p_\rR,T_\rL)$ data]{\small Energy current as a function of $p_\rR$, $T_\rL$ with fixed $p_\rL=0$ and $T_\rR=1.0$ (left, $\Delta T \ge 0$) and $T_\rR=0.3$ (right, $\Delta T \le0$). Note that $J^m\le 0$ in both cases and for all the values of $p_\rR$ considered  (details in the main text). In these and the following contour plots, the isoline corresponding to $J^e=0$ is identified by a violet full line. The uphill region in the right plot lies below the $J^e=0$ curve of approximate equation $T_\rL = \frac{\alpha}2 p_\rR^2 + T_\rR$. The current behaves as expected with respect to~$\Delta T$ ({\em i.e.} negative linear response) in both plots.}	 			
    \label{fig:Fig03} 
  \end{center}
\end{figure}

The following discussion relies on the expressions~\eqref{eq:JQ} of the thermal current~$J^Q(x)$ and~\eqref{eq:Jm}  of the mechanical current~$J^m(x)$.
Contour plots of the energy current as a function of $p_\rR,T_\rL$ at fixed values of $p_\rL,T_\rR$ are shown in Figures~\ref{fig:Fig03} and~\ref{fig:Fig05}, while Figure~\ref{fig:Fig06} shows those of $J^e(p_\rR,T_\rR)$ at fixed values of $p_\rL,T_\rL$ and Figure~\ref{fig:Fig11} shows those of $J^e(T_\rL,T_\rR)$ at fixed values of $p_\rL,p_\rR$.
\paragraph{Isolines $J^e = 0$ as a function of boundary values.}
When present, the violet full contour line in the plots identifies the isoline~$J^e=0$. The form of this curve can be found by fixing $J^e = 0$ in~\eqref{eq:19}: 
\[
-\partial_x\left( \frac{p_{\ss}^2}2\right) = \frac{\kappa(T_{\ss})}{D^{p}(T_{\ss})} \partial_x T_{\ss},
\]
and integrating both sides from $x=-1$ to~$x=1$ to obtain 
\[
\frac{p_\rL^2-p_\rR^2}{2} = \int_{T_\rL}^{T_\rR} \frac{\kappa(\theta)}{D^p(\theta)} \dd \theta.
\]
Approximating very roughly the integral on the right-hand side with a midpoint rule as 
\[
\int_{T_\rL}^{T_\rR} \frac{\kappa(\theta)}{D^p(\theta)} \dd \theta = \frac{\kappa(T^*)}{D^p(T^*)} (T_\rR-T_\rL),\qquad T^*=\frac{T_\rL+T_\rR}2,
\]
we obtain
\begin{equation}
  \label{eq:approx_Je_0}
  T_\rR-T_\rL = \frac{\alpha(T_\rL,T_\rR)}2 (p_\rL^2 - p_\rR^2), 
\end{equation}
where $\alpha(T_\rL,T_\rR) = D^p(T^*)/\kappa(T^*)>0$. The previous equation is an approximation of the level set~$J^e = 0$ for the parameters which are varied. Comparing the actual values of~$J^e$ computed by solving~\eqref{eq:19} and the level curve obtained from~\eqref{eq:approx_Je_0}, we can check ex-post that the latter curve satisfactorily approximates the level set $J^e = 0$ for the boundary values $p_\rL,p_\rR,T_\rL,T_\rR$ we consider.

\paragraph{Results for $p_\rL = 0$ and $T_\rR$ fixed.}
We first discuss the energy current results in Figure~\ref{fig:Fig03}. In both contour plots, $p_\rL=0$, $p_\rR \in [-2,2]$ and $T_\rL \in [0.3,1]$, while the right temperature~$T_\rR$ is equal to~1 in the left plot, and to~0.3 in the right one. The mechanical current $J^m$ is always nonpositive because of the boundary condition and the considered range of~$p_\rR$ values --~indeed, since $p_\rL=0$ in both cases, negative values of  $p_\rR$ imply $J^p >0 $ and $p_\ss(x)\le0$ for every $x\in[-1,1]$, while positive values of $p_\rR$ imply $J^p<0$ and $p_\ss(x) \ge0$ for every $x \in [-1,1]$. 
On the other hand, we expect the heat flow to have the same sign as $-\Delta T$, hence nonpositive in the left plot where $\Delta T\ge0$, and nonnegative in the right plot where $\Delta T \le0$. It is no surprise then, that there is no uphill current for the parameter range considered in the left plot of Figure~\ref{fig:Fig03}.
Note that the approximate equation~\eqref{eq:approx_Je_0} of the isoline $J^e=0$ in this case is $T_\rL = \frac{\alpha}2 p_\rR^2 + T_\rR$, so that uphill diffusion can occur here only for sufficiently large values of~$T_\rL$, which lie outside the considered range.
\begin{figure}
  \begin{center}
    \includegraphics[width=0.49\textwidth]{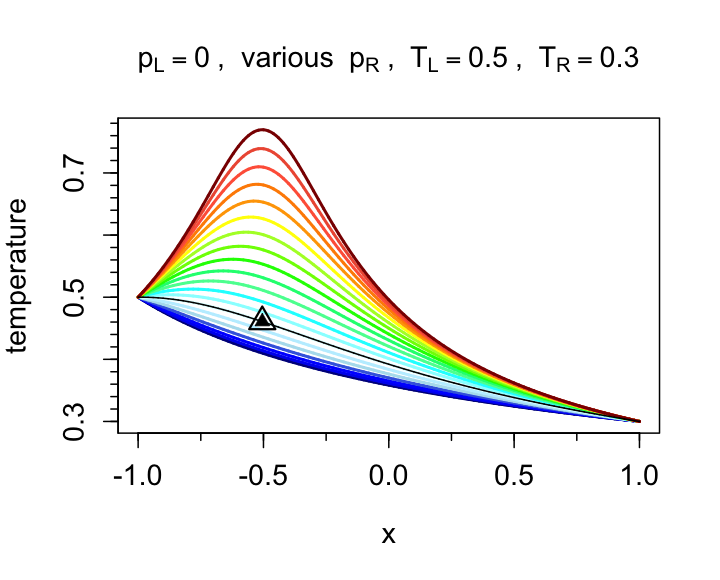}
    \includegraphics[width=0.49\textwidth]{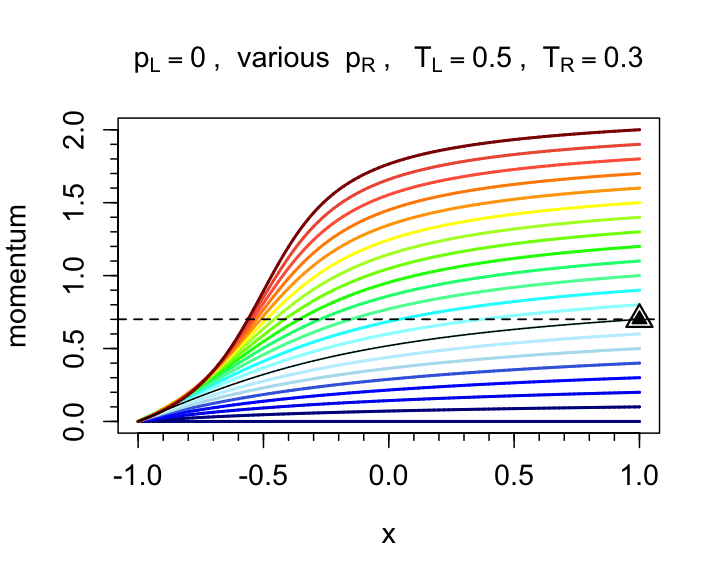}\\
    \includegraphics[width=0.49\textwidth]{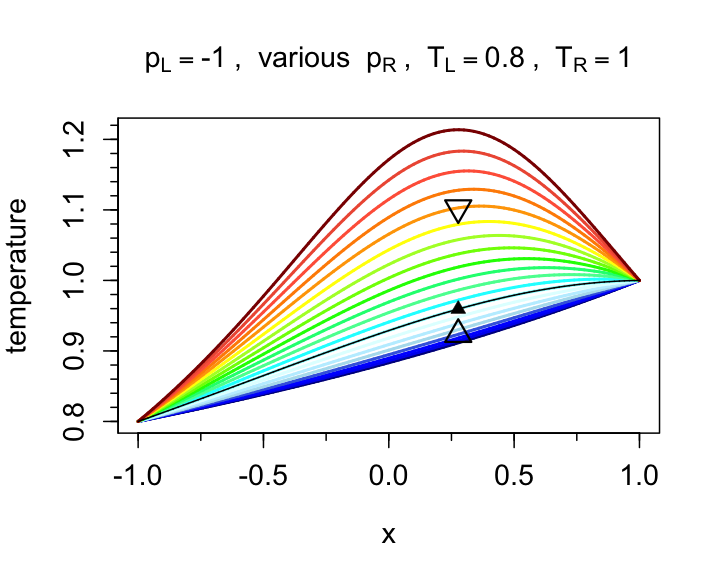}
    \includegraphics[width=0.49\textwidth]{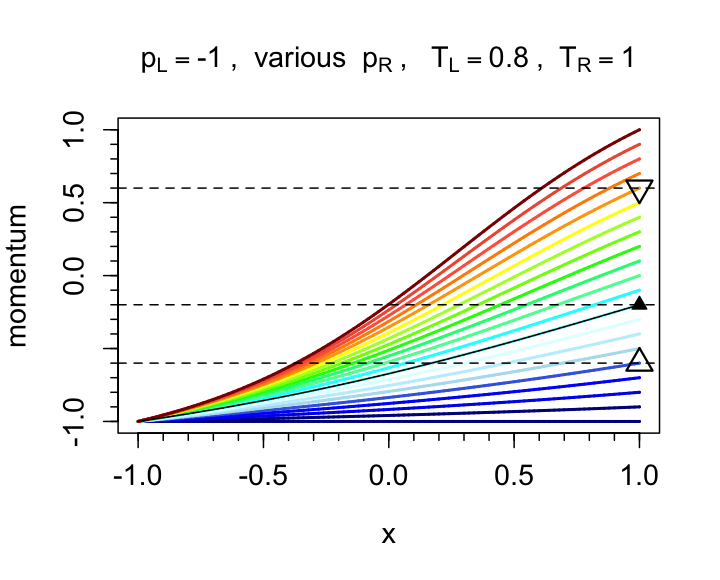}\\
    \caption[profiles data 1]{\small Profiles of temperatures (left column) and momentum (right column), for fixed $T_\rL$, $T_\rR$ and $p_\rL$ (indicated in each plot title) and values of $p_\rR$ varying from 0 (dark-blue profiles) to 2 (dark-red profiles) in steps of~0.1 on the top row, and from -1.0 (dark-blue profiles) to 1.0 (dark-red profiles) in steps of~0.1 on the bottom row. The black full-line temperature profiles labeled by the symbol ``$\blacktriangle$'' are those for which $\partial_x T_\ss \big|_{x=-1}=0$ or $\partial_x T_\ss \big|_{x=1}=0$, and correspond to the values $p_\rR=p_\rR^\blacktriangle \approx 0.7$ on the top row and to $p_\rR=p_\rR^\blacktriangle\approx -0.2$ on the bottom row (see the corresponding momentum profiles, also labeled by~``$\blacktriangle$''). For values $p_\rR >p_\rR^\blacktriangle$, the temperature profile has a maximum inside the interval $[-1,1]$. The symbols ``$\vartriangle$'' and ``$\triangledown$'' identify the curves corresponding to the values $p_\rR^\vartriangle$ and $p_\rR^\triangledown$ such that uphill diffusion arises for $p_\rR \in [p_\rR^\vartriangle,p_\rR^\triangledown]$ (see Figure~\ref{fig:Fig05}). 
   	In the bottom row plots $-p_\rR^\vartriangle=p_\rR^\triangledown\approx 0.6$, while in the top row plots, $p_\rR^\vartriangle\equiv p_\rR^\blacktriangle  \approx 0.7$ and uphill diffusion occurs for all $p_\rR \ge p_\rR^\vartriangle$.}
    \label{fig:Fig04} 
  \end{center}
\end{figure}  
The approximate isoline $J^e=0$ appears instead for the range of parameters considered in the right plot of Figure~\ref{fig:Fig03}. The uphill region defined by~\eqref{eq:UHdef} ($J^e<0$ in this case) corresponds to the region of parameters below the level set~$J^e=0$ of approximate equation $T_\rL =\frac\alpha 2 p_\rR^2 + 0.3$, that is for values of $p_\rR$ which are sufficiently large compared to the temperature difference. Furthermore, definition~\eqref{eq:UHdef} can be rewritten as
\[
\partial_x T_\ss\big|_{x=-1} > - \dfrac{p_\rL |J^p|}{\kappa(T_\rL)} = 0.
\]
This implies that, for a given value of~$T_\rL$, 
\begin{enumerate}[label=(\alph*)]
\item the onset of uphill diffusion corresponds to the value of~$p_\rR$ such that $\partial_x T_\ss\big|_{x=-1}=0$ and $J^e=0$;
\item for values of~$|p_\rR|$ greater than a threshold value (which in view of~\eqref{eq:approx_Je_0} approximately translates into $|p_\rR| > \sqrt{2(T_\rL-T_\rR)/\alpha}$\,), we have both uphill diffusion and the emergence of a maximum of $T_\ss$ inside the interval $[-1,1]$, in agreement with the discussion in Section~\ref{sec:maximum-inflection-other}. In fact, $x_{_{T_\ss^{\rm max}}} =-1$ for $J^e=0$, while $x_{_{T_\ss^{\rm max}}} \in (-1,0]$ for $J^e < 0$. 
\end{enumerate}
The upper plots in Figure~\ref{fig:Fig04} show the momentum and temperature profiles for~$p_\rL=0$, $T_\rR=0.3$ and $T_\rL=0.5$. A black triangle identifies the curve corresponding to the value $p_\rR^\blacktriangle$ for which $\partial_x  T_\ss \big|_{x=-1}=0$. 
The symbol ``$\vartriangle$'' labels the profiles corresponding to the  values~$p_\rR^\vartriangle$ such that uphill diffusion is observed for all~$p_\rR\ge p_\rR^\vartriangle$.
All profiles have the same behavior (increasing nonlinearities, appearance of a temperature peak, steepening of the momentum profiles) as those in Figure~\ref{fig:Fig02}, which were discussed at the end of Section~\ref{sec:description}.

\begin{figure}
  \begin{center}
    \includegraphics[height=0.25\textheight]{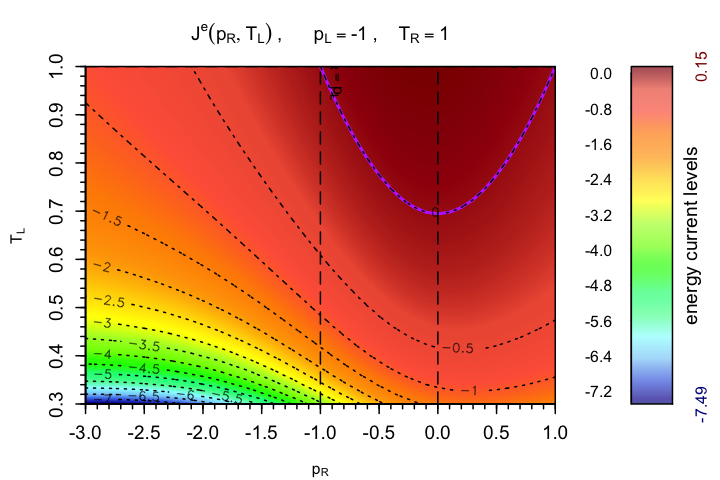}
    \includegraphics[height=0.25\textheight]{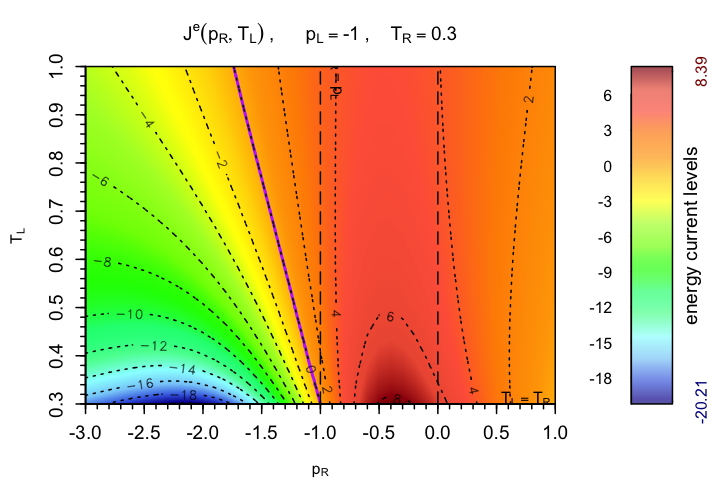}
    \caption[$J^e(p_\rR,T_\rL)$ data]{\small Energy current as a function of $p_\rR$, $T_\rL$ with fixed $p_\rL=-1$, $T_\rR=1$ (left, $\Delta T\ge 0$) and $T_\rR=0.3$ (right, $\Delta T \le 0$). In both cases the mechanical current $J^m$ is nonpositive for values of $-3\le p_\rR \le-1$ and nonnegative for $-1 < p_\rR \le 1$ (details in the main text). The violet full line represents the $J^e=0$ isoline. 
   	The uphill region lies above (below) the $J^e=0$ isoline of approximate equation $T_\rL = \frac\alpha 2 p^2_\rR + 1-\frac\alpha 2$ ($T_\rL = \frac\alpha 2 p^2_\rR + 0.3-\frac\alpha 2$) in the left (right) plot. We remark that a negative energy conductivity appears for $p_\rR$ approximately in $[-0.7,0.2]$ in the right plot (see the discussion in Section~\ref{ssec:negTC}).}	 			
    \label{fig:Fig05} 
	\end{center}
\end{figure}

\paragraph{Results for $p_\rL = -1$  and $T_\rR$ fixed.}
Figure~\ref{fig:Fig05} shows the energy current contour plots obtained by setting $p_\rL=-1$ and leaving all the other parameter values and ranges fixed as in Figure~\ref{fig:Fig03}. In both plots $J^m\le 0$ for values of $-3\le p_\rR \le-1$ (since $J^p\ge0$ and $p(x) \le 0$ for all $x\in[-1,1]$) and $J^m\ge0$ for $-1 < p_\rR \le 1$ (since $J^p \le 0$, while $p(x)<0$ for all $p_\rR \in [-1,0)$ and is ``mostly'' nonnegative for $p_\rR \in [0,1]$, see Figure~\ref{fig:Fig04}, bottom row). Note that $J^m = 0$ when $p_\rR =1$ and $T_\rL=T_\rR=1$, since the profile $p(x)$ is symmetric with respect to the origin, that is $-p(x)=p(-x)$. The ``normal'' heat flow would be nonpositive in the left plot ($\Delta T\ge 0$) and nonnegative in the right plot ($\Delta T \le 0$), thus no uphill is expected in the left half of the left plot nor in the right-half of the right plot ($p_\ss(x) \Delta p\Delta T>0$ in both cases).
The uphill diffusion region appears in the left plot of Figure~\ref{fig:Fig05} above the $J^e=0$ curve, which corresponds approximately to $T_\rL = \frac\alpha 2 p^2_\rR + 1-\frac\alpha 2$, and in the right plot below the $J^e=0$ curve, which corresponds approximately to $T_\rL = \frac\alpha 2 p^2_\rR + 0.3-\frac\alpha 2$.
The momentum and temperature profiles corresponding to the $T_\rL=0.8$ line of the left plot in Figure~\ref{fig:Fig05} are shown in the bottom row of Figure~\ref{fig:Fig04}. The symbols ``$\vartriangle$'' and ``$\triangledown$'' identify the profiles corresponding to the values~$p_\rR^\vartriangle,p_\rR^\triangledown$ such that uphill diffusion is observed for~$p_\rR\in[p_\rR^\vartriangle,p_\rR^\triangledown]$.

\begin{figure}
  \begin{center}
    \includegraphics[height=0.25\textheight]{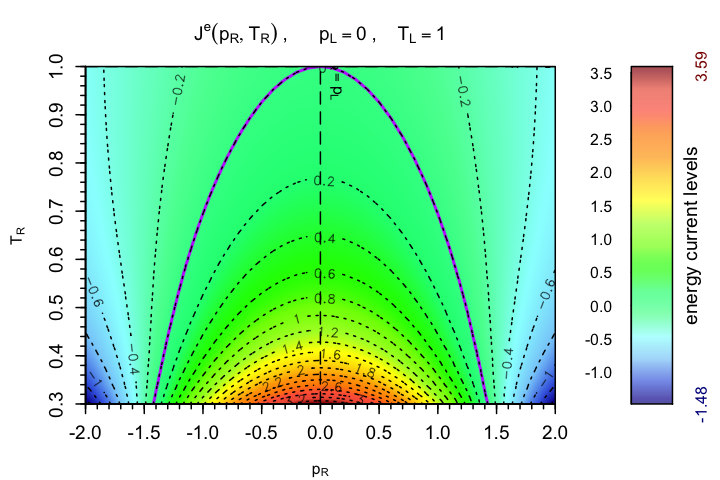}
    \includegraphics[height=0.25\textheight]{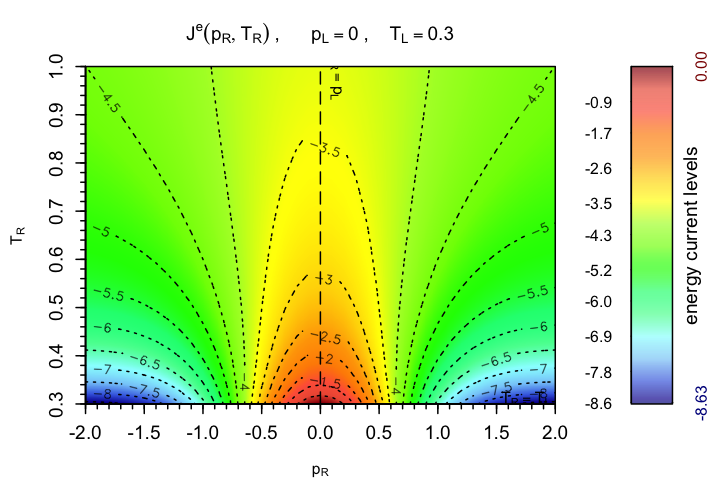}
    \caption[$J^e(p_\rR,T_\rL)$ data]{\small Energy current as a function of $p_\rR$, $T_\rR$ with fixed $p_\rL=0$, $T_\rL=0.3$ (left, $\Delta T \le0$) and $T_\rL=1.0$ (right, $\Delta T \ge 0$). Note that~$J^m\le 0$ in both cases, thus a uphill region appears only in the left plot above the $J^e=0$ isoline of approximate equation $T_\rR = 1- \frac{\alpha}2 p_\rR^2$. A negative energy conductivity appears for $|p_\rR|$ greater than approximately~1.5 in the left plot and for $|p_\rR|$ greater than approximately~0.6 in the right plot.}	 			
    \label{fig:Fig06} 
  \end{center}
\end{figure}

\paragraph{Results for $p_\rL = 0$ and $T_\rL$ fixed.}
Figure~\ref{fig:Fig06} shows the energy current contour plots for $p_\rL=0$, $T_\rL=1$ (left) and $T_\rL=0.3$ (right). In both cases $J^m \le 0$ as in Figure~\ref{fig:Fig03}, while the expected heat flow is nonnegative in the left plot ($\Delta T \le 0$) and nonpositive in the right plot ($\Delta T \ge 0$). Therefore, the conditions for uphill energy diffusion are only met in the $T_\rL=1$ case for the parameter range we consider. The equation of the isoline $J^e=0$ is approximately $T_\rR = 1- \frac{\alpha}2 p_\rR^2$, thus concave, in agreement with the values of~$J^e$ reported in the plot. Since $T_\rL \ge T_\rR$, the uphill region is above this curve.  

\begin{figure}
  \begin{center}
    \includegraphics[height=0.3\textheight]{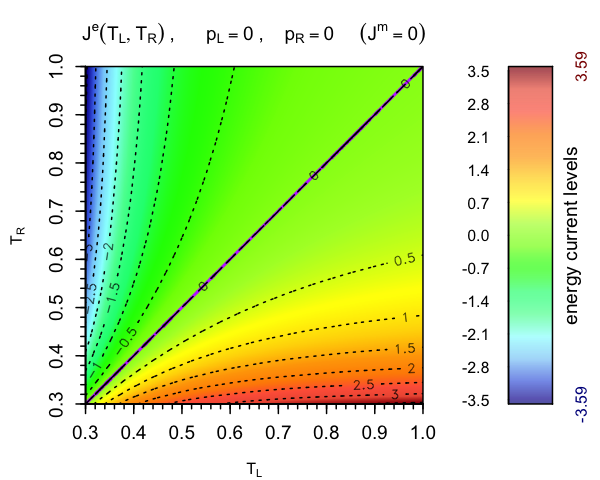}
    \includegraphics[height=0.3\textheight]{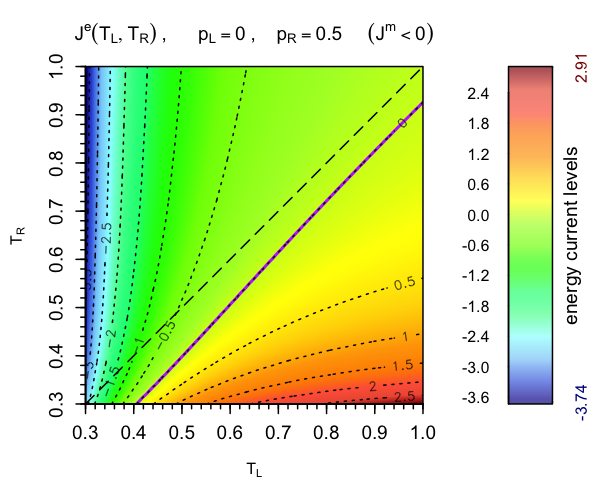}\\
    \includegraphics[height=0.3\textheight]{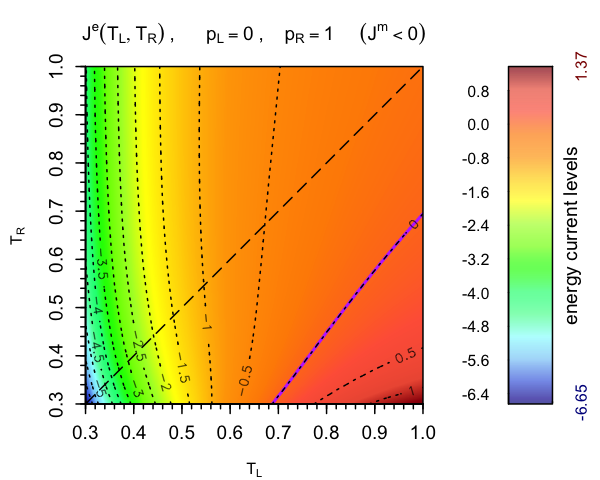}
    \includegraphics[height=0.3\textheight]{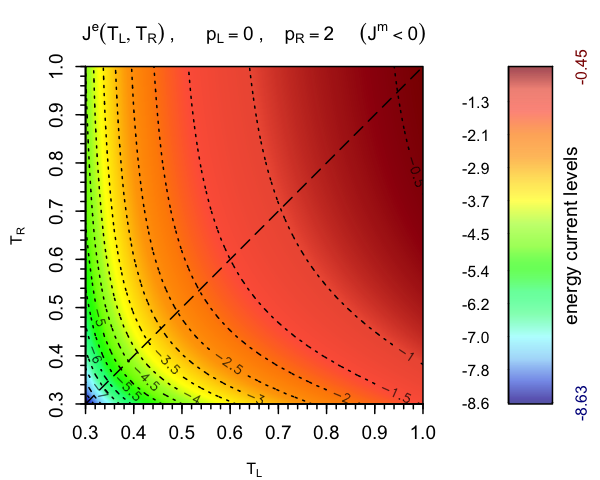}
    \caption[$J^e(p_\rR,T_\rL)$ data]{\small Energy current as a function of $T_\rL$ and $T_\rR$ with fixed $p_\rL=0$, $p_\rR=0$ (top left, $J^m =0$), $p_\rR=0.5$ (top right, $J^m < 0$), $p_\rR=1$ (bottom left, $J^m <0$) and $p_\rR=2$ (bottom right, $J^m <0$). In all plots, $\Delta T$ is positive above the diagonal line and negative below, so the energy current is expected to be negative above and positive below that line. {\em Uphill\/}: In the top right and bottom left plots, an uphill region is present and lies between the $J^e=0$ isoline (approximately $T_\rR=T_\rL - \frac\alpha 2$) and the $T_\rR=T_\rL$ line, while in the bottom right plot, the uphill region lies below the diagonal $T_\rR=T_\rL$. {\em Negative energy conductivity\/}: The top left plot shows the "normal conductivity" context. In the absence of mechanical forcing,  $J^e$ has the sign of~$-\Delta T$ and is monotonic with respect to~$-\Delta T$. At $p_\rR = 0.5$ (top right plot) the response of the system is still ``normal''. The bottom left plot shows that some negative energy conductivity emerges for $p_\rR=1$ and $T_\rL$ approximately below~$0.6$, since $J^e$ increases with $T_\rR$, while it is expected to decrease. In the bottom right plot, we observe negative energy conductivity at all values of $T_\rL$. Remark that the response of the system to a variation of $T_\rL$ at fixed $T_\rR$ is normal in all four plots.}
    \label{fig:Fig11} 
  \end{center}
\end{figure}

\paragraph{Results for $p_\rL = 0$ and $p_\rR$ fixed.}
Figure~\ref{fig:Fig11} shows some contour plots of the energy current as a function of~$T_\rL$ and~$T_\rR$. 
We observe that in this case the $J^e=0$ curve is a straight line, which is consistent with~\eqref{eq:approx_Je_0}. The top-left plot shows the results for $\Delta p=0$, and the isoline $J^e=0$ corresponds to~$T_\rL=T_\rR$, thus to equilibrium conditions. There is no uphill diffusion in this case since there is no mechanical current. The top-right plot displays the behavior of $J^e$ for~$p_\rR =0.5$ as a function of the boundary temperatures. An uphill diffusion region appears between the diagonal $T_\rR=T_\rL$ and the $J^e=0$ isoline. In the bottom-left plot, the isoline $J^e=0$ corresponds approximately to $T_\rR=T_\rL - \frac\alpha 2$, and the area of the uphill diffusion region has increased. In the bottom right plot, the region of parameters associated with uphill diffusion is that below the diagonal $T_\rR=T_\rL$. 

\subsection{Negative energy conductivity}
\label{ssec:negTC}

\begin{figure}
  \begin{center}
    \includegraphics[width=0.49\textwidth]{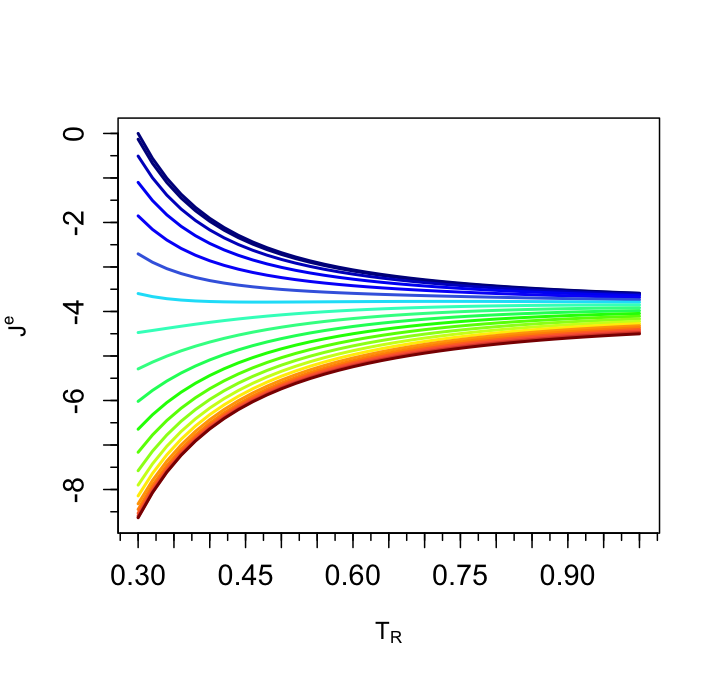}
    \includegraphics[width=0.49\textwidth]{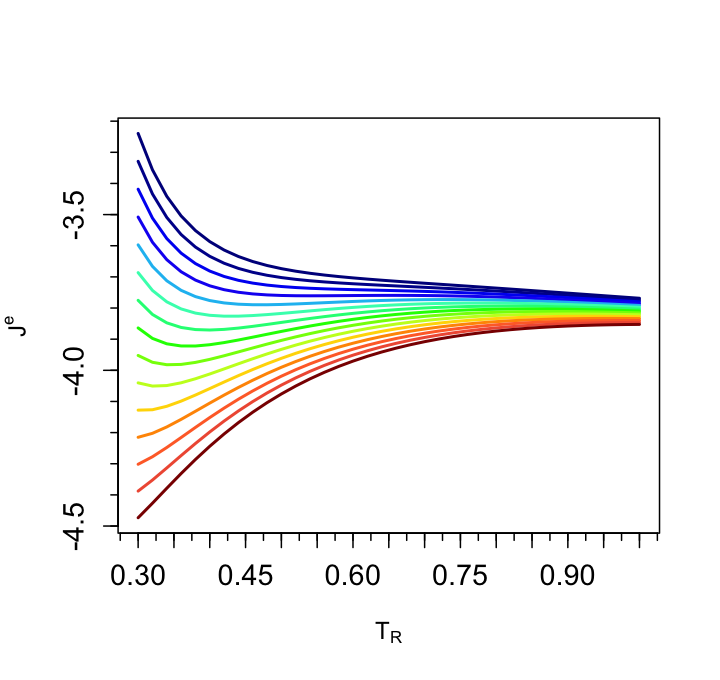}\\
    \caption[currents 1D]{\small Energy current $J^e$ as a function of $T_\rR$ at $T_\rL=0.3$. Left: $J^e(T_\rR)$ for different values of $p_\rR$, starting from $p_\rR=0$ (dark-blue line) to $p_\rR=2$ (dark-red line) in steps of~0.1. Right: zoom on the range of values $p_\rR \in [0.56,0.7]$ in steps of~0.01. These plots show how the response of the system to an increase of~$\Delta T = T_\rR - T_\rL \ge 0$ depends on the value of~$p_\rR$. In the right panel, we clearly see the emergence of a minimum of~$J^e$ for which $\dd J^e/\dd T_\rR =0$ for values of $p_\rR \in (0.6,0.66)$. In these cases, the energy conductivity is normal up to a temperature~$T_\rR$ corresponding to the minimum of $J^e$, and becomes negative at larger temperatures.} 			
    \label{fig:Fig07} 
  \end{center}
\end{figure} 
Thanks to the macroscopic dynamics equations~\eqref{eq:19} we were able to reproduce the results in~\cite{ILOS11} and extensively study the occurrence of negative energy conductivity for a wide range of parameters. The most striking case is that shown in Figure~\ref{fig:Fig06}, right plot. On the vertical line $p_\rR=0$, since $\Delta p=0$, we observe the normal behavior of the energy current: $J^e$ is negative and decreases as $T_\rR$ increases (so that $\left|J^e\right|$ increases). 
As soon as $p_\rR \ne 0$, a negative mechanical current emerges, which has the expected effect of globally decreasing~$J^e$. This can be observed more easily in Figure~\ref{fig:Fig07},
where we report the behavior of $J^e(T_\rR)$ at $T_\rL=0.3$ for various values of $|p_\rR|$. 
However, the response of the system to the increase of $T_\rR$ gradually changes as $|p_\rR|$ increases: $J^e$ decreases with $T_\rR$ for values of $\left|p_\rR\right|$ below approximately~0.6 (normal conductivity), and it becomes an increasing function of~$T_\rR$ for $|p_\rR|$ greater than approximately~0.66. To the latter regime corresponds the remarkable phenomenon of negative energy conductivity, following the definition in Section~\ref{sec:neg-cond-diff}. There is an intermediate regime for values of $\left|p_\rR\right|$ approximately between~0.6 and~0.66 (Figure~\ref{fig:Fig07}, right plot), where~$J^e$ is a non-monotonic function of~$T_\rR$ and admits some global minimum for a finite value of~$T_\rR > T_\rL$. In this intermediate regime, the energy conductivity is normal for sufficiently small values of~$T_\rR$, and becomes negative at some larger value of $T_R$.

\begin{figure}
  \begin{center}
    \includegraphics[width=0.49\textwidth]{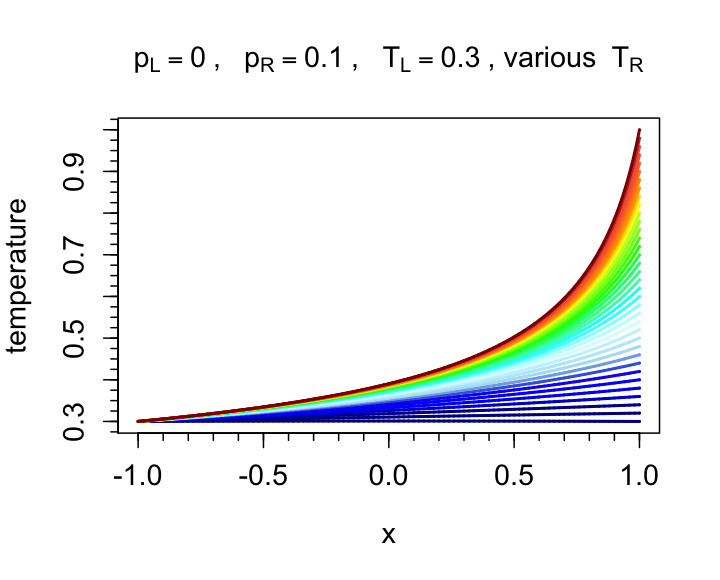}\hspace{-2mm}
    \includegraphics[width=0.49\textwidth]{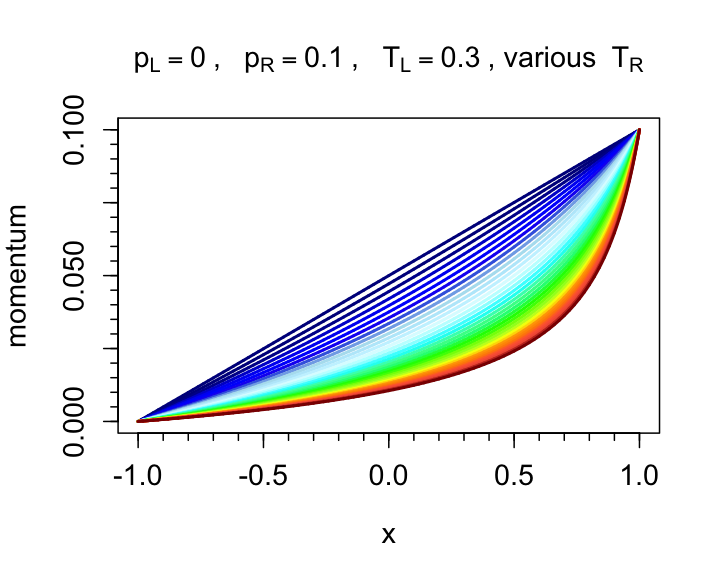}\\
    \includegraphics[width=0.49\textwidth]{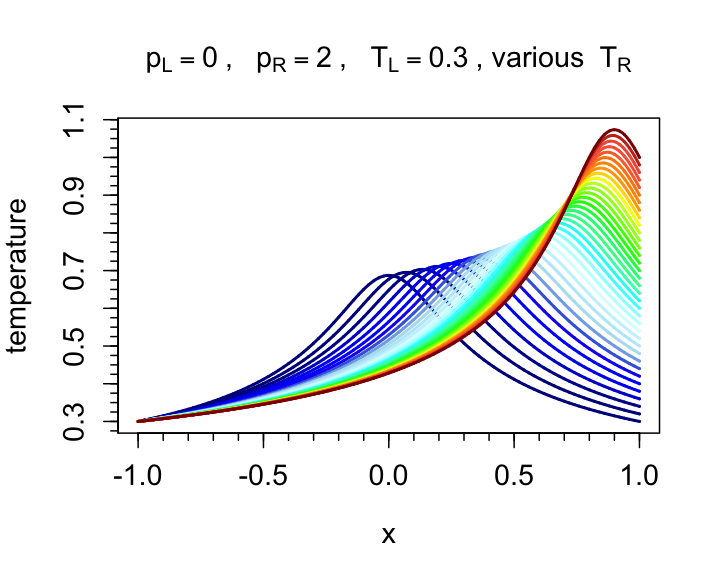}\hspace{-2mm}
    \includegraphics[width=0.49\textwidth]{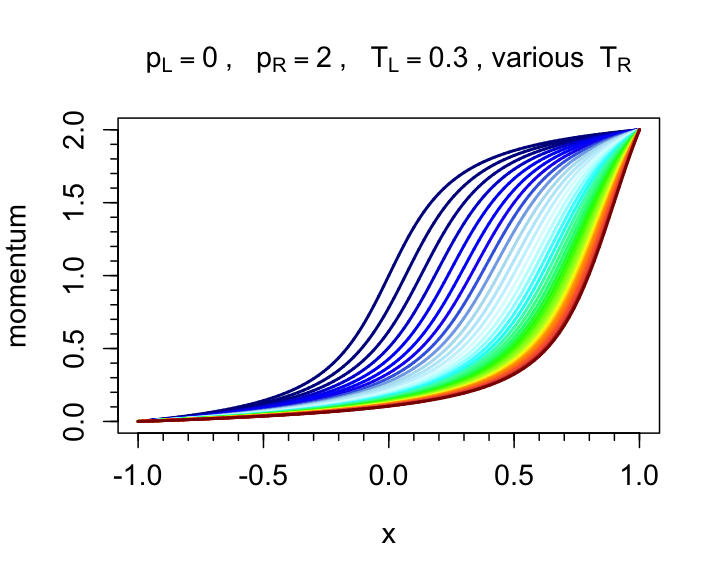}\\
    \caption[profiles data]{\small Profiles of temperatures (left column) and momentum (right column), for fixed $p_\rL=0$, $T_\rL=0.3$, $p_\rR$ (indicated in each plot title) and values of $T_\rR$ varying from 0.3 (dark-blue profiles) to 1.0 (dark-red profiles) in steps of~0.1. The top plots correspond to $p_\rR=0.1$ which is contained in the normal conductivity range, while the bottom plots correspond to $p_\rR=2$ which lies in the negative energy conductivity range (see Figure~\ref{fig:Fig06}). Note in particular the different behaviors of $\partial_x T_\ss \big|_{x=-1}$ as $T_\rR$ increases: this derivative is increasing for small values of~$p_\rR$ (top left) and decreasing for large values of $p_\rR$ (bottom left).}	 			
    \label{fig:Fig08} 
  \end{center}
\end{figure}

\begin{figure}
  \begin{center}
    \includegraphics[width=0.45\textwidth]{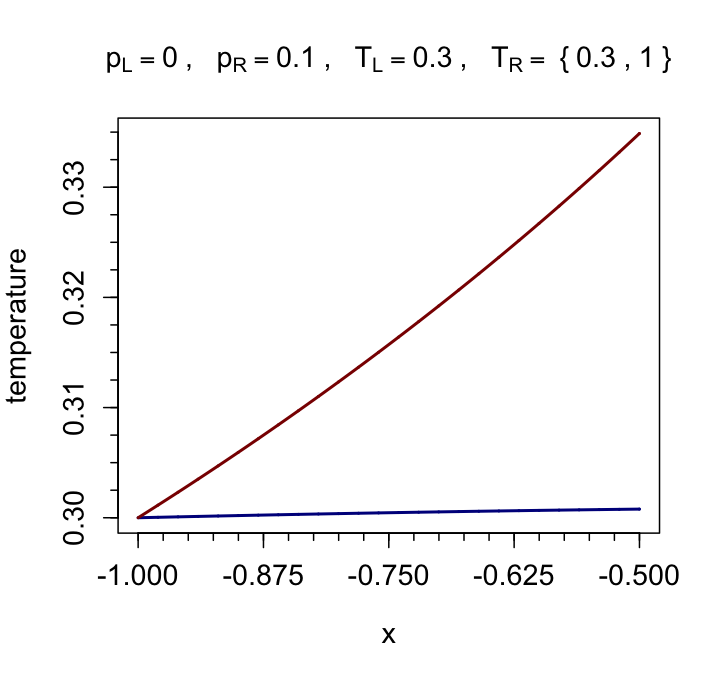}\hspace{-2mm}
    \includegraphics[width=0.45\textwidth]{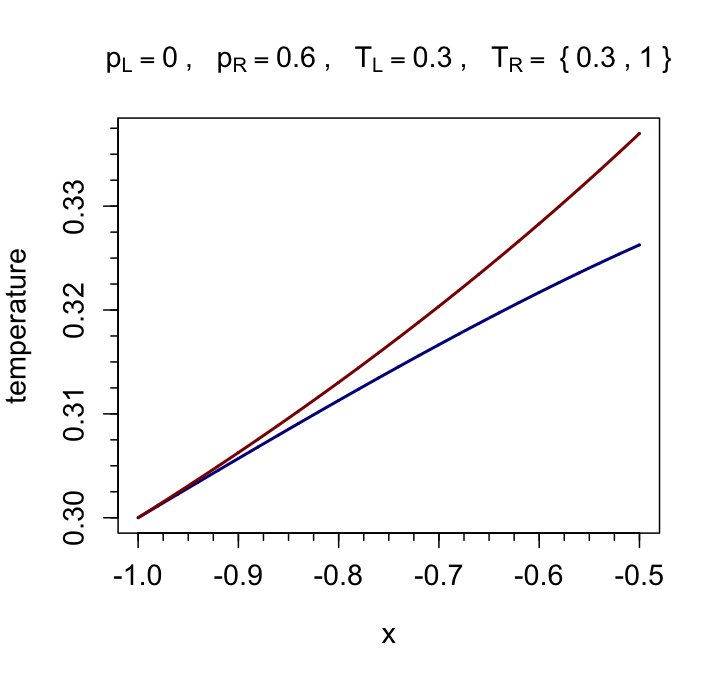}\\
    \includegraphics[width=0.45\textwidth]{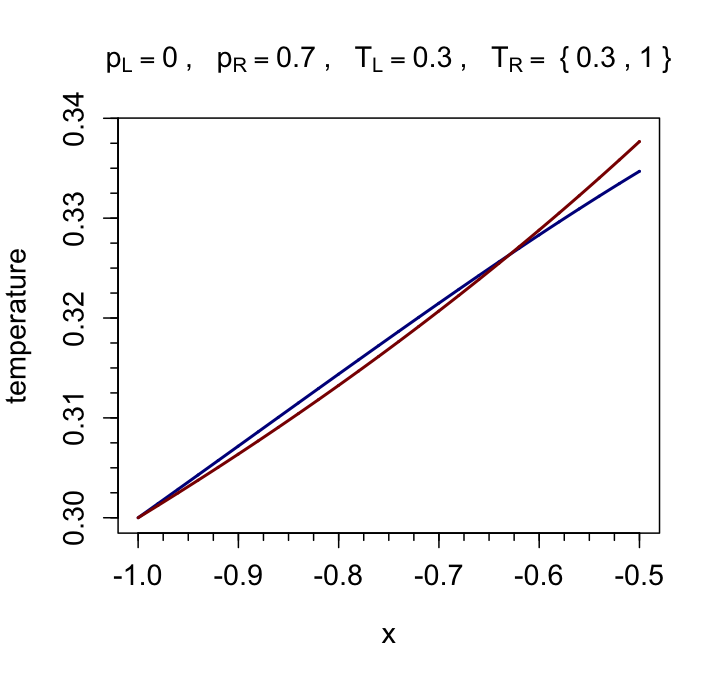}\hspace{-2mm}
    \includegraphics[width=0.45\textwidth]{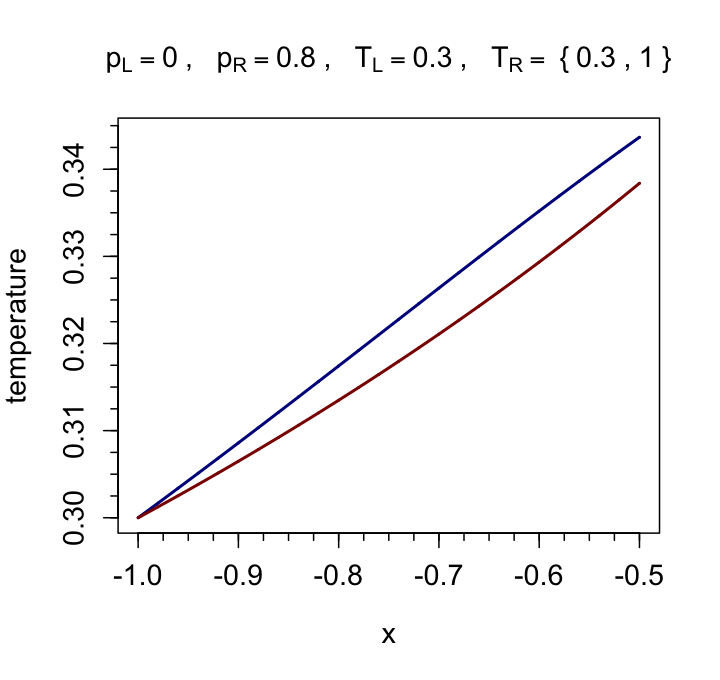}\\
    \caption[profiles data detail]{Detail of the temperature profiles plotted in Figure~\ref{fig:Fig08} for $T_\rR=0.3$ (dark-blue line) and $T_\rR = 1.0$ (dark-red line), and $p_\rR$ equal to 0.1 (top left), 0.6 (top right), 0.7 (bottom left) and 0.8 (bottom right).}
    \label{fig:Fig09} 
  \end{center}
\end{figure} 
The corresponding temperature and momentum profiles are plotted in Figure~\ref{fig:Fig08} for $p_\rR=0.1$ (top row, ``normal conductivity'') 
and $p_\rR=2.0$ (bottom row, ``negative conductivity''), displaying a striking change in the behavior of $\partial_x T_\ss\big|_{x=-1}$ with respect to the increase of the right boundary temperature: it increases when the value of $|p_\rR|$ is small and decreases when the latter is large. This behavior is emphasized in Figure~\ref{fig:Fig09}, where we plot the details of the temperature profiles relative to $T_\rR=0.3$  and $T_\rR=1.0$ in the interval $x\in[-1,-0.5]$, for $p_\rR = \{0.1,0.6,0.7,0.8\}$.
For values of $\left|p_\rR\right|$ below approximately~0.6, we observe that $\partial_x T_\ss\big|_{x=-1}$ increases as $T_\rR$ increases, while for $\left|p_\rR\right|$ above approximately~0.7, $\partial_x T_\ss\big|_{x=-1}$ decreases as $T_\rR$ increases. Thus, in view of~\eqref{eq:condizia}, which rewrites here for $p_\rL=0$ as
\[
J^e = -\kappa(T_\rL) \partial_x T_\ss\big|_{x=-1},
\]
we obtain that $J^e$ increases (\emph{i.e.}~becomes less negative) as $T_\rR$ increases for values of $\left|p_\rR\right|$ approximately greater than~0.7, which leads to a negative energy conductivity. 
Note that an increase of the temperature at the right boundary leads to a decrease of the temperature close to the left boundary, which is another somewhat counter-intuitive phenomenon.

\begin{figure}[ht]
  \begin{center}
    \includegraphics[width=0.5\textwidth]{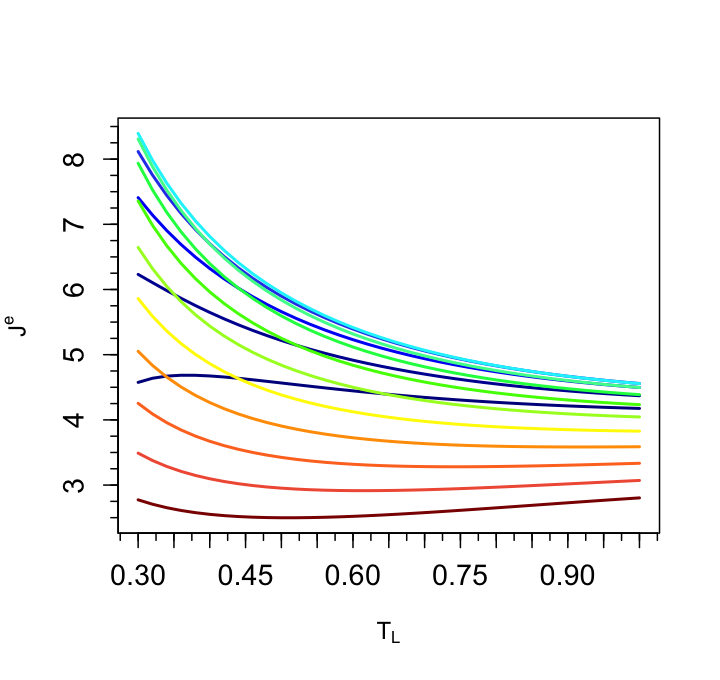}
    \caption[energy current]{Energy current at fixed $p_\rL=-1.0$, $T_\rR=0.3$, for  $p_\rR$ from a value of~$-0.8$ (dark-blue line) to~$0.5$ (dark-red line). This plot shows how the response of the system to an increase of $\Delta T\ge 0$ varies with $p_\rR$.}	 	
    \label{fig:Fig10} 
  \end{center}
\end{figure}

\begin{figure}
  \begin{center}
    \includegraphics[width=0.49\textwidth]{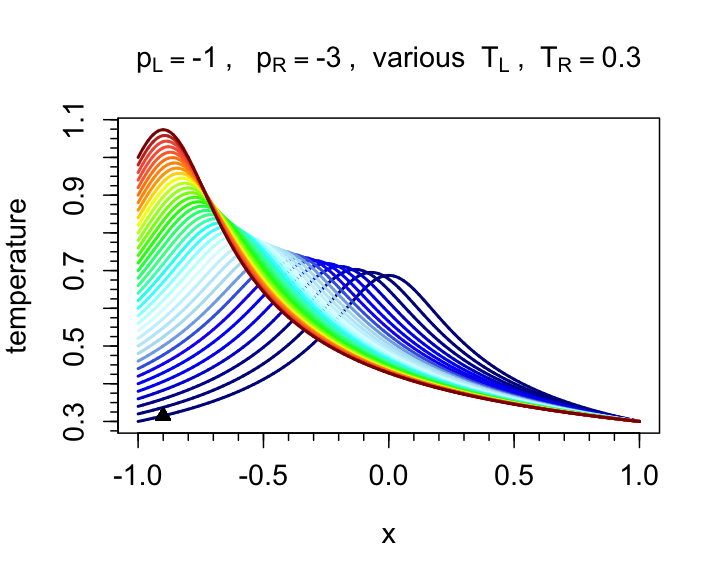}\hspace{-2mm}
    \includegraphics[width=0.49\textwidth]{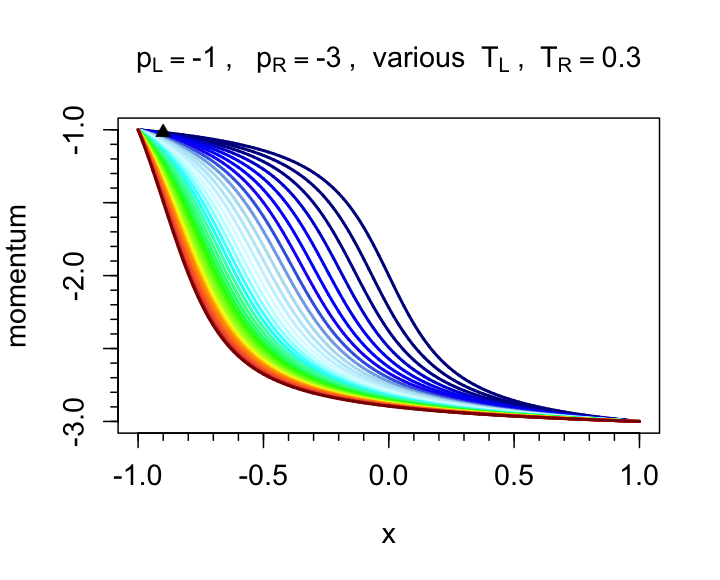}\\
    \includegraphics[width=0.49\textwidth]{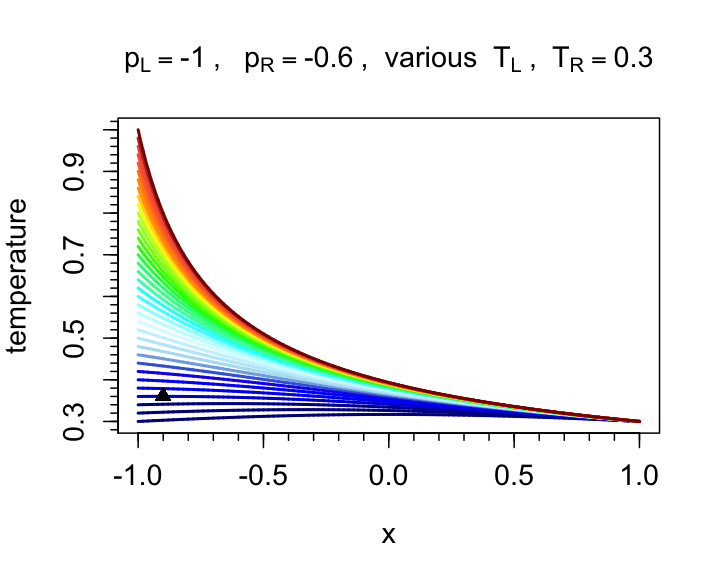}\hspace{-2mm}
    \includegraphics[width=0.49\textwidth]{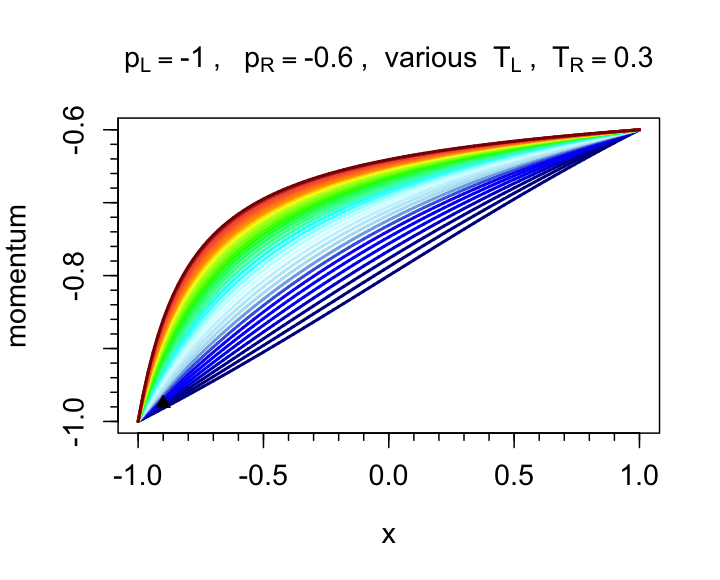}\\
    \caption[profiles data]{\small Profiles of temperatures (left column) and momentum (right column), for fixed $p_\rL=-1.0$, $T_\rR=0.3$, $p_\rR$ (indicated in each plot title) and values of $T_\rL$ varying from 0.3 (dark-blue profiles) to 1.0 (dark-red profiles) in steps of~0.1. The top plots correspond to a value of $p_\rR=-3.0$ which is in the normal conductivity range, while in the bottom plots the value of $p_\rR=-0.6$ is included in the negative energy conductivity range (see also Figure~\ref{fig:Fig05}, right plot).}
    \label{fig:Fig14}
  \end{center}
\end{figure}

A negative energy conductivity region also appears in the left plot of Figure~\ref{fig:Fig06}. In this case $J^e$ should decrease as $T_\rR$ increases. Instead, for $|p_\rR|$ approximately greater than~1.5, $J^e$ (slightly) increases, as can be seen from the contour lines. Another case of negative energy conductivity is shown in Figure~\ref{fig:Fig05} (right plot). In this case, at fixed $p_\rL\le p_\rR$, $J^e$ should increase as $T_\rL$ increases. Instead, it decreases for $p_\rR$ approximately contained in the interval $[-0.7,0.2]$, as shown in Figure~\ref{fig:Fig10}.
The temperature and momentum profiles for the latter set-up, with~$p_\rR=-3$ and $p_\rR=-0.6$, are displayed in Figure~\ref{fig:Fig14}. In contrast to the situation considered in Figure~\ref{fig:Fig08}, there is no evident mechanism that can explain through~\eqref{eq:condizia} why $J^e$ decreases as $\Delta T$ is decreased.

In Figure~\ref{fig:Fig11}, the negative energy conductivity can be seen from a different perspective. Here, the ``normal conductivity'' behavior is observed in the top plots, where $J^e$ increases (resp. decreases) as $\Delta T$ is decreased (resp. increased). When $p_\rR >0$, the response for $T_\rR<T_\rL$ changes: for fixed $T_\rL$ (approximately below~0.6 in the bottom left plot, and in the range~$[0.3,1.0]$ in the bottom right plot), $J^e$ increases as $T_\rR$ increases. We remark that the response of the system to a variation of $T_\rL$ at fixed $T_\rR$ is always normal.

The numerical evidence gathered by our simulations indicates that we can observe a negative response of the system when the boundary value of the momentum is different from zero and the boundary temperature is modified on the same side. This is true in all the situations in which we vary $T_\rR$ with $p_\rR \neq 0$, and when we vary $T_\rL$ with $p_\rL=-1.0$.
Note that negative energy conductivity can arise both when $p_\ss(x) \Delta p\Delta T>0$ (see Figure~\ref{fig:Fig05}, right; Figure~\ref{fig:Fig06}, right) and when $p_\ss(x) \Delta p\Delta T<0$, which is the necessary condition for uphill diffusion (see Figure~\ref{fig:Fig06}, left). In particular, in Figure~\ref{fig:Fig11} we have both phenomena. 
	
Besides the qualitative observations collected in this section, we are unable to give precise conditions or general explanations of the phenomenon of negative energy conductivity. However, we hope that our approach and numerical results will trigger further research on this topic.

\appendix

\section{Computation of the Green-Kubo coefficients}
\label{App:GKnum}

We first discuss how to obtain estimates of~\eqref{eq:9}--\eqref{eq:9bis} by microscopic simulations, and then how to fit~$\kappa,D^p$ on these data. 

\paragraph{Estimation of~\eqref{eq:9}--\eqref{eq:9bis} by microscopic simulations.}
In order to numerically approximate~\eqref{eq:9}--\eqref{eq:9bis}, we introduce several discretization parameters:
\begin{enumerate}[label=(\roman*)]
\item the size~$M$ of the finite systems considered in the numerical integration;
\item the number~$R$ of realizations used to approximate the expectation with an empirical average;
\item the time step~$\Delta t>0$ used in the integration of the dynamics and the discretization of the time integral;
\item the largest number of time steps~$L$ over which the dynamics is integrated (which corresponds to truncating the time integral to the upper bound~$L\Delta t$).
\end{enumerate}
We consider rotor chains of size~$M$ with periodic boundary conditions. Realizations are obtained by sampling initial conditions $q^r = (q_1^r,\dots,q_M^r)$ and $p^r = (p_1^r,\dots,p_M^r)$ according to the canonical distribution at the desired temperature ($1 \leq r \leq R$, $r$ being the index of the realization), and then numerically integrating in time the Hamiltonian dynamics with the standard Verlet scheme~\cite{Verlet}. Currents at the $\ell$-th time step for the $r$-th realization are denoted by~$J^a_{r,\ell}$ for $a \in \{e,p\}$. The discrete approximation of~\eqref{eq:9}--\eqref{eq:9bis} is then, 
\begin{equation}
  \label{eq:numGKcoeff}
  \widehat K^{a,a} = \dfrac{\Delta t}R \sum_{\ell=1}^{L} \sum_{r=1}^R \widetilde{J}^a_{r,0} \widetilde{J}^a_{r,\ell}\,,\qquad \widetilde{J}^a_{r,\ell} = J^a_{r,\ell} - \frac 1R \sum_{r'=1}^R J^a_{r',\ell}\,, \qquad J^a_{r,\ell} = \frac 1M\sum_{i=1}^M j^{a,r,i}_{r,\ell},
\end{equation}
where $j_{i,i+1}^{a,r,\ell}$ is the instantaneous local $a$-current flowing between sites~$i$ and~$i+1$ at time $\ell\Delta t$ for replica~$r$. In practice, we consider $\Delta t =10^{-2}$. 
Note that we compute an empirical covariance of the current by centering the numerical approximations of the average currents (which may be not centered due to biases caused by the time step and the number of realizations). 

To generate the initial conditions, we start from $q^r_i = 0$ for $1 \leq i \leq M$ and $p^r_i$ sampled from the Gaussian distribution whose variance is the target temperature. The system is then thermalized for a time~50 by a Langevin dynamics with a fluctuation/dissipation of magnitude~$\gamma=1$ acting on all momenta. We check that the marginal distributions of~$(r,p)$ at the end of this thermalization procedure are indeed canonical distributions at the desired target temperature. 

The parameters $M$, $L$ and $R$ must be chosen large enough for the time covariance of the energy-energy and momentum-momentum currents to decay to zero. We find that for $M \ge 500$ the results do not essentially vary with respect to the parameter~$M$, so we fix it to~$500$ in all our simulations. We also considered $R=10^6$ realizations. As for the number of iterations, we chose $L\Delta t = 12,000$ for $T=0.1$, $L\Delta t=10,000$ for $T=0.15$, $L\Delta t=2,000$ for $0.2\le T < 0.3$, $L\Delta t=1,000$ for $0.3 < T < 0.5$ , $L\Delta t=300$ for $0.5 \le T \le 0.7$, $L\Delta t=150$ for $0.7 \le T <1$ and $L\Delta t=100$ for $1 \le T \le 3 $.

\paragraph{Numerical fit of the data.}
The values of $\widehat K^{a,a}$ are numerically fitted by the following functions
\begin{equation}
\widehat K^{p,p}\left(T^{-1}\right) = a_{1}  \rme^{-b_{1} T} +\dfrac{c_1}{T^2}, \qquad
\widehat K^{e,e}\left(T^{-1}\right) = a_{2} + \dfrac{b_{2}}{T} +\dfrac{c_2}{T^2},
\label{eq:fitGKcoeff}
\end{equation}  
where $a_1 = -5.00 \pm 1.32$, $b_1=2.11 \pm 0.23$, $c_1=0.95 \pm 0.06$ and $a_2=0.20\pm 0.02$, $b_2=0.20\pm 0.03$,
$c_2=0.176 \pm 0.008$. We have chosen the functional forms in~\eqref{eq:fitGKcoeff} after a meticulous analysis among other classes of functions as the best performing in the range of temperatures~$[0.3,1.5]$. This range roughly corresponds to the range of temperatures~$T_\ss$ observed in the numerical integration of the macroscopic differential system~\eqref{eq:19} (see Section~\ref{sec:numerics}). We checked that there were no significant qualitative changes in our results when considering other functional forms for the functions in~\eqref{eq:fitGKcoeff}.

\section{Computation of the stationary solution of the 
macroscopic\texorpdfstring{\newline}\phantom{e}quations}
\label{App:SSnum}

We present in this section the numerical method used to solve~\eqref{eq:19} with the boundary conditions~\eqref{eq:bc}. The unknowns are the profiles $p_\ss$, $T_\ss$, as well as the values of the currents $J^p,J^e$. We rely on a fixed-point strategy where we first update the currents, then the momentum profile, then the temperature profile. From~\eqref{eq:24} and~\eqref{eq:25}, we obtain the expressions of the currents in terms of the profiles
\begin{equation}
  J^p = - \frac{p_\rR - p_\rL}{I_1}, \qquad J^e = - \frac{T_\rR-T_\rL}{I_2} + \frac{I_3}{I_2} J^p,
\label{eq:constJpandJe}
\end{equation}
with
\begin{equation}
\label{eq:integrals_Ix}
  I_1 = \int_{-1}^1 \frac{1}{D^p(T_\ss(x))} \, \dd x, \qquad I_2 = \int_{-1}^1 \frac{1}{\kappa(T_\ss(x))} \, \dd x, \qquad I_3 = \int_{-1}^1 \frac{p_\ss(x)}{\kappa(T_\ss(x))} \, \dd x.
\end{equation}

In order to discretize the profiles $p_\ss,T_\ss$, we introduce a mesh of the domain~$[-1,1]$ with internal nodes~$x_k = -1+k\Delta x$ for $1 \leq k \leq K$, where $(K+1)\Delta x = 2$. The approximation of the values of $p_\ss(x_k)$ and $T_\ss(x_k)$ are denoted by $p_k$ and $T_k$, respectively. We consider the following discrete counterpart of~\eqref{eq:19}
\begin{equation}
\begin{aligned}
J^p_{\Delta x} &= - \frac 12 \left(D^p(T_k) + D^p(T_{k+1})\right) \dfrac{p_{k+1} - p_k}{\Delta x}\,, \\
J^e_{\Delta x} &= - \frac 12 \big(\kappa(T_{k}) + \kappa(T_{k+1}) \big) \dfrac{T_{k+1} - T_k}{\Delta x} +\frac 12\, \big(p_{k} + p_{k+1}\big)\, J^p_{\Delta x}\,, 
\end{aligned}
\qquad 1 \leq k \leq K\,,
\label{eq:discrete_eq}
\end{equation}
together with the boundary conditions $p_0 = p_\rL$, $p_{K+1} = p_\rR$, $T_0 = T_\rL$ and $T_{K+1} = T_\rR$. Note that the values of the currents depend in principle on~$\Delta x$.

The initial profiles $p^{0} = \{ p_k^{0} \}_{1 \leq k \leq K}$ and $T^{0} = \{ T_k^{0} \}_{1 \leq k \leq K}$ are obtained by a linear interpolation between the fixed boundary values. For a given tolerance $\varepsilon > 0$, the algorithm inductively construct updates $p^n = \{ p_k^n \}_{1 \leq k \leq K}$ and $T^n = \{ T^n_k \}_{1 \leq k \leq K}$ of the profiles as follows: for $n\ge 0$,
\begin{enumerate}[label=(\arabic*)]
\item Compute approximations $I_{1,\Delta x}^{n}$, $I_{2,\Delta x}^{n}$ and $I_{3,\Delta x}^{n}$ of the integrals in~\eqref{eq:integrals_Ix} from the profiles~$T^n$ and~$p^n$ via~\eqref{eq:fitGKcoeff} with a Simpson's quadrature rule; 
\item Update the values of the currents using~\eqref{eq:constJpandJe}:
  \[
    J^{p,n}_{\Delta x} = - \frac{p_\rR - p_\rL}{I_1^{n}}, \qquad
    J^{e,n}_{\Delta x} = - \frac{T_\rR-T_\rL}{I_2^{n}} + \frac{I_3^{n}}{I_2^{n}} J^{p,n}_{\Delta x}; 
  \]
\item Update the momentum profile from the first of~\eqref{eq:discrete_eq} as $\widetilde p_0^{n+1} = p_\rL$ and, for $k = 1,\dots,K+1$,
  \[
    \widetilde p_{k+1}^{n+1} = \widetilde{p}_{k}^{n+1} - \frac{2 \Delta x}{D^p(T_{k+1}^n)+ D^p(T_{k}^n)} J^{p,n}_{\Delta x};
  \]
\item Update the temperature profile from the second of~\eqref{eq:discrete_eq} as $\widetilde T_0^{n+1} = T_\rR$,  and, for $k = 1,\dots,K+1$,
      \[
	\widetilde T_{k+1}^{n+1} = \widetilde{T}_{k}^{n+1} - \frac{2 \Delta x}{\kappa(T_{k+1}^n)+ \kappa(T_k^n)} \left(J^{e,n}_{\Delta x} - \frac{\widetilde{p}_{k+1}^{n+1}+\widetilde{p}_{k}^{n+1}}{2} J^{p,n}_{\Delta x}\right);
      \]
\item Define the new profiles using a mixing rule with parameters $\alpha_p$ and $\alpha_T$: for $1 \leq k \leq K+1$,
\begin{align*}
p_{k}^{n+1} & = \widetilde p_{k}^{n+1} + \alpha_p (p_{k}^{n}- \widetilde p_{k}^{n+1}),\\
T_{k}^{n+1} & = \widetilde T_{k}^{n+1} + \alpha_T (T_{k}^{n}- \widetilde T_{k}^{n+1});
\end{align*}
\item If both $\dps \left\|p^{n+1} - p^n\right\|_\infty := \max_{k=0,\dots,K}\left|p^{n+1}_k - p_k^n\right| \leq \varepsilon$ and $\left\|T^{n+1}-T^n\right\|_\infty \leq \varepsilon$, stop; otherwise increase~$n$ by~$1$ and go back to Step~(1).
\end{enumerate}

Note that a sign of a ``good'' convergence of the algorithm at a cycle $n=\bar n$ is that $p_{K+1}^{\bar n+1} \approx p_\rR$ and $T_{K+1}^{\bar n+1} \approx T_\rR$.

For the simulations reported in this work, we chose $\varepsilon = 1.5 \times 10^{-8}$ and $\Delta x = 0.002$. Concerning the mixing parameters~$\alpha_p$ and~$\alpha_T$, we fixed their value to~0.9 after a series of tests on the worst convergence cases, namely those with the lowest value of the boundary temperatures and the highest values of~$\Delta p$.

\paragraph{Acknowledgments.}
The authors thank Christophe Poquet and Guillaume Legendre for fruitful discussions. This work was partially supported by the ANR-15-CE40-0020-01 grant LSD.

\bibliographystyle{plain} 
\bibliography{IOS_RotorsNL-REV}

\end{document}